\documentstyle[12pt,epsfig]{article}

\def\lapprox{\lower .7ex\hbox{$\;\stackrel{\textstyle <}{\sim}\;$}}
\def\d{{\rm d}}
\def\calN{{\cal N}}
\begin{document}

\begin{titlepage}
\vspace*{-1cm}
\begin{flushright}
DTP/95/26 \\
March 1995 \\
(Revised May 1995)
\end{flushright}
\vskip 1.cm
\begin{center}
{\Large\bf Deep Inelastic Electron--Pomeron \\[2mm]
Scattering at HERA}
\vskip 1.cm
{\large T.~Gehrmann} and {\large W.J.~Stirling}
\vskip .4cm
{\it Department of Physics, University of Durham \\
Durham DH1 3LE, England }\\
\vskip 1cm
\end{center}
\begin{abstract}
The idea that the pomeron has partonic structure similar
to any other hadron has been given strong support by recent
measurements of the diffractive structure function at HERA.
We present a detailed theoretical analysis of the diffractive
structure function under the assumption that
the diffractive cross section can be factorized into
a pomeron emission factor and the deep inelastic scattering cross
section of the pomeron.
We pay particular attention to the kinematic
correlations implied by this picture, and suggest the measurement of
an angular correlation which should provide a first test of the
whole picture.
We also present two simple phenomenological
models for the quark and gluon structure
of the pomeron, which are consistent with various theoretical ideas and
which give equally good fits to recent measurements by the H1 collaboration,
when combined with the pomeron emission factor of Donnachie and
Landshoff. We predict that a large fraction of diffractive deep
inelastic events will contain charm, and discuss how improved data
will be able to distinguish the models.
\end{abstract}
\vfill
\end{titlepage}
\newpage

\section{Introduction}
\label{sec:ds}
Recent measurements at HERA have indicated that  a significant fraction
of deep inelastic electron-proton scattering  events
have  a final state with a
large rapidity gap between the proton beam direction and the
observed final state particles \cite{zeus,h1}.
The lack of any hadronic activity
around the proton beam direction
and the mismatch between the initial-state and observed final-state
energy requires the proton (deflected only
by a small angle and therefore outside the rapidity coverage of the
detectors) to be in the final state, still carrying a large fraction of
its incident momentum. These events with a remnant proton in the
final state are classified as diffractive scattering (DS).
 In analogy to
the conventional deep inelastic scattering (DIS) cross section,
the differential cross section for DS can be written
as
\begin{equation}
\label{diffX}
\frac{\d\sigma^{D\!S}}{\d\alpha_{p}\d t
\d x \d Q^2} \;  =
\; \frac{4\pi\alpha^2}{xQ^4}\; \left\{
1-y+\frac{y^2}{2[1+R^{D\!S}(x,Q^2,\alpha_p,t)]}\right\} \; F_2^{D\!S}
(x,Q^2,\alpha_p,t).
\end{equation}

\begin{figure}
\begin{center}
{}~ \epsfig{file=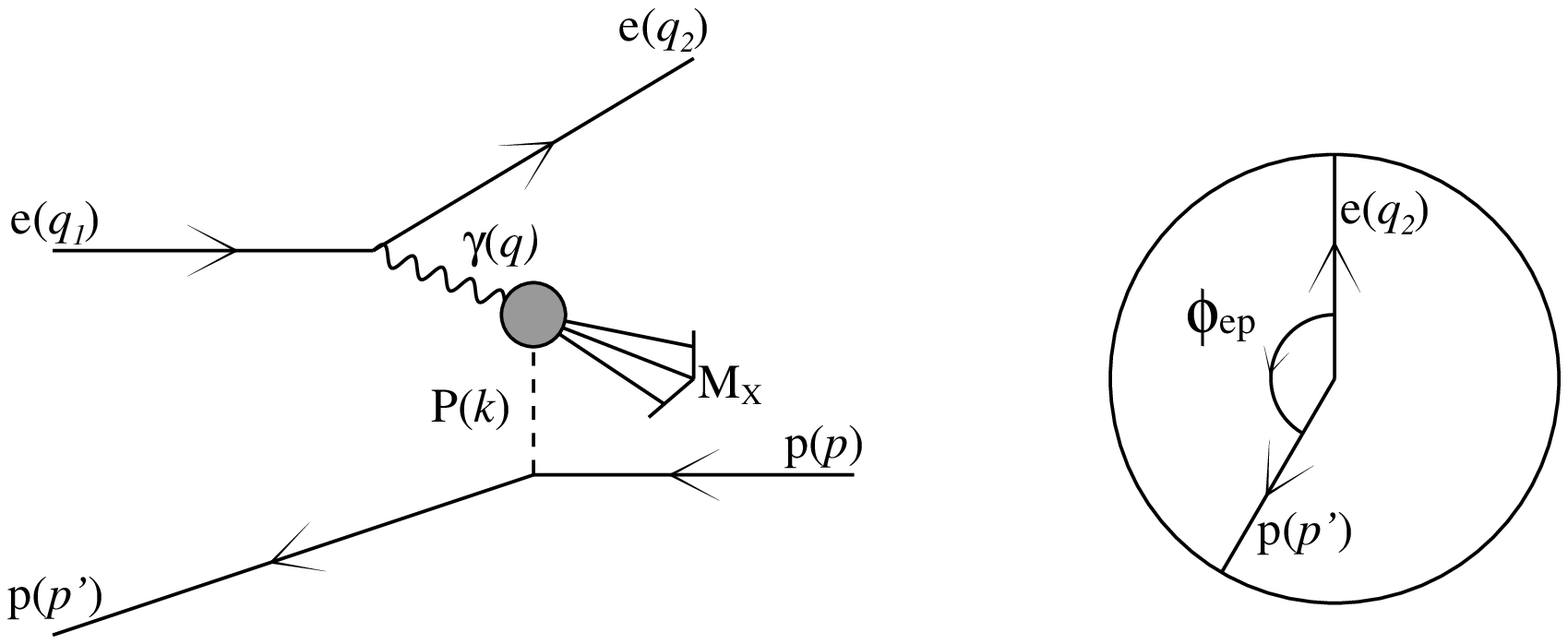,width=13cm}
\caption{Kinematics of deep inelastic electron-pomeron scattering}
\label{fig:one}
\end{center}
\end{figure}

The final state configuration of these events suggests that they are
caused by a deep inelastic scattering of an uncharged and uncoloured
object, which was emitted from the proton
beforehand, Fig.~\ref{fig:one}. From the
kinematical distribution of the diffractive events it seems most
likely that this object is the pomeron, which  so far has only been
observed and identified by its $t$-channel trajectory \cite{dldata} in
the full
hadron-hadron cross section.\footnote{There has been recent evidence
\protect{\cite{WA91}}
for a glueball candidate at $M=1.9$ GeV, which lies on the timelike
continuation of the pomeron trajectory and has the correct quantum
numbers predicted by Regge theory.}
The idea that the pomeron has hard partonic constituents was first
proposed by Ingelman and Schlein  \cite{ing1},
 and given strong support by the hadron collider experiments of the UA8
collaboration \cite{UA8,SCHLEIN}.

If this interpretation is correct, then one would expect that the
diffractive cross section could be factorized into a piece
corresponding to the emission of an uncharged,
colourless pomeron from the proton and a piece corresponding to a hard
scattering off the partonic constituents of the pomeron:
\begin{equation}
\label{facexact}
\frac{\d\sigma^{D\!S}}{\d\alpha_{p}\d t
\d z \d Q^2} \;  =
\; \frac{4\pi\alpha^2}{zQ^4}\; \left\{
1-y_P+\frac{y_P^2}{2[1+R^{P}(z,Q^2,t)]}\right\} \; F_2^{P}
(z,Q^2,t) f(\alpha_p,t),
\end{equation}
where $z=z(x,Q^2,\alpha_p,t)$ is the fraction of pomeron momentum
carried by the struck parton and
\begin{equation}
y_P=\frac{yx}{\alpha_p z}
\label{yexact}
\end{equation}
 is the virtual photon energy `seen' by the pomeron.
$F_2^{P}(z,Q^2,t)$ denotes the
DIS structure function of the pomeron and
$f(\alpha_{p},t)$ the probability for the emission of a pomeron
with momentum fraction $\alpha_{p}$ and $t$-channel momentum $t$ off a
proton.

A common but
only approximately correct way of parametrizing this factorization
property is to write the diffractive structure
function as the product of an
emission factor and the deep inelastic structure function of the pomeron
\cite{dldis,ing2}:
\begin{equation}
\label{f2ds}
F_2^{D\!S}(x,Q^2,\alpha_p,t) = F_2^{P}(z,Q^2,t)\; f(\alpha_{p},t).
\end{equation}
Note that  the $R^{D\!S}$ and $R^{P}$ functions
cannot be related in such a simple
manner. We will discuss various tests of the factorizability of the
cross section and investigate the applicability of the factorization
at the level of structure functions (\ref{f2ds}).

Due to the lack of experimental information on the remnant proton, a
complete kinematical reconstruction of diffractive scattering
events is not possible at present. Both parameters describing the
pomeron emission ($\alpha_p$ and $t$) can only estimated indirectly
or have to be integrated out.
The kinematical parameter $z$, describing the fraction of
the pomeron's light-cone momentum `seen' by the virtual photon,
can be (up to a small uncertainty) obtained by measuring the invariant
mass of the hadronic system $X$ in Fig.~\ref{fig:one}. Since to a good
approximation $ \alpha_p = x/z$,  the $\alpha_p$ dependence
of $F_2^{D\!S}$ can be inferred and compared with theoretical predictions
for the function $f$. In this approxmiation, the factorizability of
the structure function (\ref{f2ds}) becomes an exact statement
following from the factorizability of the cross section
(\ref{facexact}).
We will discuss the reconstruction of the kinematics
and the uncertainty on $\alpha_p$ caused by the lack of
kinematical information on the remnant proton in Section~\ref{sec:kin}.

If  the object struck by the virtual photon in diffractive
deep inelastic scattering
is indeed the same pomeron which controls the high energy behaviour
of hadronic scattering amplitudes, then its basic properties
and in particular its coupling to the proton are already known.
For example,
Donnachie and Landshoff give a simple form for
$f(\alpha_{p},t)$ \cite{dldis} which they derive from an essentially
nonperturbative\footnote{There have been several recent
attempts to derive a perturbative formulation of the pomeron. These
approaches \cite{bfklph}, all based on the BFKL equation \cite{bfkl}, will
not be discussed in the context of this paper, as there is
insufficient  conclusive evidence at present for the
applicability of the BFKL equation in
the kinematic range covered at HERA. In the following discussion, we
will always assume $f(\alpha_p,t)$ (as for any other
hadron-hadron interaction at low invariant momentum transfer) to
represent a nonperturbative
coupling of pomerons to the proton, which can be determined from the
experimental data.}
model \cite{dlpom} of pomeron
exchange dynamics in terms of Regge amplitudes:
\begin{equation}
\label{fpom}
f(\alpha_p,t) = \frac{9b^2}{4\pi^2}\left[ F_1(t)\right]^2
\alpha_p^{1-2\alpha(t)}.
\end{equation}
The Dirac form factor of the proton entering in $f(\alpha_p,t)$ is
well known from low-energy $ep$ scattering \cite{dirac}:
\begin{equation}
\label{formfac}
F_1(t)= \frac{4 M^2 - 2.8 t}{4 M^2 - t} \left( \frac{1}
{1-t/(0.7\; {\rm GeV}^2)}\right)^2,
\end{equation}
where $M$ is the proton mass, whereas the pomeron coupling
strength to quarks\footnote{We use the notation $b$ rather
than $\beta$ to avoid confusion with the kinematic variable
introduced in Section~\ref{sec:data}.}
$b\approx 1.8\ {\rm  GeV}^{-1}$ and the pomeron
trajectory
\begin{equation}
\label{traj}
\alpha(t) = 1 + \epsilon + \alpha ' t \quad \mbox{with} \quad \epsilon
= 0.086, \quad \alpha ' = 0.25\ {\rm GeV}^{-2}
\end{equation}
are  tuned to explain a wide range of experimental results
in elastic $pp$, $p\bar{p}$, and $\pi p$ scattering \cite{dldata}.
Other similar forms  for $f$ have been proposed in the literature,
see for example Ref.~\cite{ing1},
but the differences are not crucial to the present discussion.

The above picture has recently been given strong support by a detailed
analysis of diffractive deep inelastic scattering events
by the H1 collaboration at HERA \cite{h1new}.
Their principle conclusions are: (i) the $Q^2$ dependence of
$F_2^{D\!S}$ is consistent with scattering off pointlike objects,
 (ii)  the factorization of the
diffractive structure function into pieces which depend
separately on $z$ and $\alpha_p$, Eq.~(\ref{f2ds}), is  observed,  (iii)
the $\alpha_p$ dependence of $f$ is consistent with that
predicted by Donnachie and Landshoff, i.e. $ \sim \alpha_p^{1-2\alpha(0)}$,
and (iv) the pomeron structure function $F_2^{P}$
 is `hard', i.e. the  pointlike constituents carry a
  significant fraction of the pomeron's momentum on average.
{\it Not} yet determined experimentally are: (i) the `nature' of
these hard constituents (i.e. whether the pomeron predominantly
consists of quarks or of gluons), (ii) the explicit $t$-dependence of
$F_2^{D\!S}$ predicted by Eqs.~(\ref{fpom},\ref{formfac},\ref{traj}),
(iii) the pomeron flux factor (i.e. a possible overall normalization factor
in Eq.~(\ref{fpom})),
(iv) the kinematical distribution of
the remnant protons, and (v) the magnitude of the $R$-factor (its impact
on the H1 results has been shown to be less than 17\% \cite{h1new}).

It is these latter issues that are the subject of the present study.
Having established experimentally
that the overall picture is consistent with Fig.~\ref{fig:one},
the next task is to ask more detailed questions.
We have already mentioned that the kinematic variable $\alpha_p$
cannot at present be measured directly, and so it is important
to study  the uncertainty which is introduced
when it is reconstructed from observed quantities.
The kinematical constraints  implicit in Fig.~\ref{fig:one} also lead to
small but non-negligible correlations between the final state
electron and proton. The magnitude of these correlations depends
on the form of $f$ and $F_2^P$. Finally we shall investigate
the $z$ and $Q^2$ dependence of $F_2^P$ itself. The H1 data already
contain a significant amount of information.

 In particular, we shall show
that the data can be understood {\it either} in terms
of a model in which  the pomeron
predominantly consists of gluons at a scale $Q_0^2=2 \,\mbox{GeV}^2$,
with a small admixture of quarks such that a momentum sum rule is
satisfied, provided
that the pomeron flux factor is rescaled, {\it or} in terms of a model
in which the pomeron, like the photon, has pointlike couplings to quarks
such that the structure function has an additional `direct' contribution.
In the case of the latter model, the pomeron flux factor requires
no further rescaling. We shall discuss in the some detail
the different $z$ and $Q^2$ behaviours expected in the two approaches.
A common feature is the prediction of a sizeable charm quark contribution
to diffractive deep inelastic scattering.

The paper is organized as follows.
In the following section we study the kinematics of diffractive
deep inelastic
scattering, as implied by Fig.~\ref{fig:one}, in some detail.
In Section~\ref{sec:corr} we present predictions for a particular
kinematic correlation which should be straightforward to measure
and which will provide a stringent test of the pomeron picture.
In Section~\ref{sec:qcd} we discuss models for the parton structure
of the pomeron, and the corresponding predictions for the $z$ and
$Q^2$ dependence of the pomeron structure function $F_2^P$.
Our predictions
are compared with the experimental data from H1 in Section~\ref{sec:data}.
Finally, Section~\ref{sec:conc} contains our conclusions.

\section{Kinematics of electron-pomeron deep inelastic scattering}
\label{sec:kin}

\subsection{Reconstruction of the kinematical invariants}
To reconstruct all kinematical parameters in (\ref{facexact}), it is
sufficient to measure the momenta of the outgoing electron ($q_2$) and the
remnant proton ($p'$).
It is convenient to parametrize these in a Sudakov decomposition using
two lightlike vectors directed along the beam and a spacelike
transverse vector. Since we are ignoring the electron mass we can use
the incoming electron momentum $q_1$ for one of the lightlike vectors.
For the other, we define
\begin{equation}
\bar p  = p - {M^2 \over s - M^2}\; q_1
\end{equation}
where $ s = (p+q_1)^2$, $p^2 = M^2$ and, by construction, $\bar p^2 = 0$.
Hence we can write
\begin{eqnarray}
q_2 &=& A \bar p + B q_1 + \vec{q_T} \nonumber \\
p' &= &C \bar p + D q_1 + \vec{k_T} ,
\label{eq:finalmom}
\end{eqnarray}
which implies
\begin{eqnarray}
q & =& - A \bar p + (1-B) q_1 - \vec{q_T} \nonumber \\
k &= &(1-C) \bar p + \left( {M^2 \over s - M^2} -D\right) q_1 - \vec{k_T} .
\end{eqnarray}
The eight degrees of freedom in (\ref{eq:finalmom}) are
reduced to five by requiring that $q_2^2 = 0$, $p'^2 = M^2$ and disregarding
an overall azimuthal angle.
The next step is to relate these to more familiar deep inelastic
and diffractive variables.
The electron is described by the usual two DIS
variables $x$ and $Q^2$, and three additional parameters
define the proton:
\begin{eqnarray}
\alpha_p & = & \mbox{fraction of longitudinal momentum transferred to
the pomeron,}\\
t & = & \mbox{$t$-channel invariant momentum transfer to
the pomeron,}\\
\phi_{ep} & = & \mbox{angle between  the outgoing electron and
outgoing proton}\nonumber\\
& & \mbox{in the plane transverse to the beam direction.}
\end{eqnarray}
In terms of Lorentz invariants,
\begin{equation}
Q^2 = -q^2, \quad x = {Q^2 \over 2 p\cdot q},\quad
t = k^2, \quad  \alpha_p  = {k\cdot q_1 \over p \cdot q_1}.
\label{invariants}
\end{equation}
Some straightforward algebra then gives the result for the
photon and pomeron momenta:
\begin{eqnarray}
q & =&  \left( {Q^2 \over s - M^2}\right)\;  \left[ - \bar p +
\left({1\over x}+ {M^2 \over s - M^2}\right) q_1 \right]
- \vec{q_T} \nonumber \\
k &=& \alpha_p \bar p +  {t - \alpha_p M^2 \over s - M^2}  q_1
 - \vec{k_T} ,
 \label{qandk}
\end{eqnarray}
where
\begin{eqnarray}
q_T^2 & = & Q^2 \left( 1 - {Q^2 \over x (s-M^2) } -
{M^2 Q^2 \over (s-M^2)^2 } \right)    \nonumber \\
k_T^2 & = &  -t (1-\alpha_p) - \alpha_p^2 M^2   \nonumber \\
\cos \phi_{ep} &= & {\vec{q_T} \cdot \vec{k_T} \over
\sqrt{ q_T^2 k_T^2 } } .
\end{eqnarray}

As already mentioned, neither $\alpha_p$ or $t$ are directly measured.
An additional constraint can however be obtained by measuring
the mass of the final state  in the  $\gamma^{\star}(q)
P(k)\rightarrow X$ hard scattering, i.e. $M_X^2 = (q+k)^2$.
In analogy with the usual Bjorken $x$ variable, Eq.~(\ref{invariants}),
 we introduce
\begin{equation}
z  =  \frac{Q^2}{2q\cdot k}
 =  \frac{Q^2}{M_X^2+Q^2-t} .
\end{equation}
Substituting the expressions for $q$ and $k$ from Eq.~(\ref{qandk})
then gives the required constraint:
\begin{eqnarray}
{1\over z}  & = & {\alpha_p\over x}\; + \; {2 \alpha_p M^2 -t
\over  s-M^2 }  \nonumber \\
& & -\; {2 \cos \phi_{ep}\over Q }\; \left[ \left\{
-t(1-\alpha_p) - \alpha_p^2 M^2\right\}\;
\left\{ 1 - {Q^2 \over x (s-M^2) } -
{M^2 Q^2 \over (s-M^2)^2 }    \right\}  \right]^{1\over 2} .
\label{master}
\end{eqnarray}

{}From Eq.~(\ref{formfac}) we see that large values  of $|t|$ are expected
to be heavily suppressed, and this is consistent with the fact that no
final-state protons are observed outside the beam pipe. It is therefore
a reasonable first approximation to set $t=0$ in the kinematical
relations. With $t=M^2=0$, Eq.~(\ref{master}) becomes
\begin{equation}
{1\over z}   =  {\alpha_p\over x} \quad \Rightarrow \quad
\alpha_p = {x\over z} ,
\label{simple}
\end{equation}
with the interpretation that the momentum fraction
of the quark in the proton ($x$)
is simply the product of the momentum fraction of the quark in the
pomeron ($z$) and the momentum fraction of the pomeron in the proton
($\alpha_p$). Note that in this approximation
\begin{equation}
\label{zsimple}
 z = Q^2/(Q^2 + M_X^2).
\end{equation}
In this way, the parameter $\alpha_p$ is easily determined from measured
quantities.

It is important to stress, however, that the  corrections
to (\ref{simple}) are not obviously negligible. In particular,
we note that the terms of order $\sqrt{-t}/Q$  and $M/Q$
may not be small, while corrections to (\ref{zsimple}) start at order
$t/(M_X^2+Q^2)$ and will therefore be ignored in the following.
In practice,
the necessity to have a large rapidity gap in order to distinguish
the diffractive events requires the pomeron to be slowly moving in the
laboratory frame, and consequently $\alpha_p \ll 1$.
In this limit, including the most important subleading corrections gives
\begin{equation}
\alpha_p = {x\over z} \;  \left[ 1 + 2 \sqrt{1-\frac{Q^2}{xs}}
\; \frac{z\sqrt{-t}}{Q}\;  \cos\phi_{ep} \right] .
\label{alpha}
\end{equation}
This result shows that the distribution in the angle $\phi_{ep}$
will not be uniform in general. For any non-zero $t$, and at fixed
$z$, $x$ and $Q^2$, $\alpha_p$ varies with $\phi_{ep}$. Since the
diffractive structure function is a steeply falling function of
$\alpha_p$, Eqs.~(\ref{f2ds},\ref{fpom}), the effect can be quite large.
This effect will be studied in greater detail below,
and in particular the
implications for angular correlations between the outgoing electron
and the remnant proton will be elaborated in Section~\ref{sec:corr}.

\subsection{Estimates for the systematic uncertainties at HERA}

The dependence of $\alpha_p$ on the presently unmeasurable angle
$\phi_{ep}$ gives rise to a systematic uncertainty
on reconstructing the variables $y_P$ and $\alpha_p$
which appear in (\ref{facexact}). In this section we attempt to quantify
these uncertainties in order to test the validity of the approximations
\begin{equation}
\label{h1appr}
\alpha_p \approx \frac{x}{z}, \qquad y_P \approx y
\end{equation}
used to extract $F_2^P(z,Q^2)$ from the HERA
data~\cite{h1new}\footnote{We use $\sqrt{s}=296\;\mbox{GeV}$ for
all following numerical evaluations.}. We will
also test the factorizability of the diffractive structure function
(\ref{f2ds}).

For any parametrization of $f(\alpha_p,t)$ which has a similar
$\alpha_p$ dependence to (\ref{fpom}), one obtains a  diffractive cross
section which decreases steeply with $\alpha_p$. Therefore the
correction (\ref{alpha}) will {\it not} average out over all angles
$\phi_{ep}$, rather these corrections will accumulate to
give a non-zero average
deviation from (\ref{h1appr}).
The relative deviation of $\alpha_p$ from $x/z$ is given by
\begin{equation}
\frac{\alpha_p-x/z}{x/z} = 2 \sqrt{1-\frac{Q^2}{xs}}\frac{z\sqrt{-t}}{Q}\cos
\phi_{ep},
\end{equation}
while the relative deviation of $y_P$ from $y$ has a similar form:
\begin{equation}
\frac{y_P-y}{y} = \frac{x}{z\alpha_p} - 1 \simeq -2
\sqrt{1-\frac{Q^2}{xs}}\frac{z\sqrt{-t}}{Q}\cos \phi_{ep} =
-\frac{\alpha_p-x/z}{x/z}.
\end{equation}
Since  $\phi_{ep}$ and $t$ are not directly measured,
we define the expectation value of
the deviation to be the weighted average over all angles and values of
$t$\footnote{We assume here that $F_2^P$ is independent of
$t$, i.e. that $f(\alpha_p,t)$ in (\ref{facexact}) takes account of
the full $t$-dependence.}:
\begin{eqnarray}
\left\langle \frac{\alpha_p-x/z}{x/z}\right\rangle (x,z,Q^2) & = &
\frac{\displaystyle \int_{-\infty}^0 \d t \int_{0}^{2\pi} \d \phi_{ep}
\left(\frac{\alpha_p - x/z}{x/z}\right) f(\alpha_p,t) }{\displaystyle
\int_{-\infty}^0 \d t
\int_{0}^{2\pi} \d \phi_{ep} f(\alpha_p,t) },\\
\left\langle \frac{y_P-y}{y}\right\rangle (x,z,Q^2) & = & -\left\langle
\frac{\alpha_p-x/z}{x/z}\right\rangle (x,z,Q^2)
\end{eqnarray}
which becomes, for any $f(\alpha_p,t)$ with a similar form to (\ref{fpom}),
\begin{equation}
\left\langle \frac{\alpha_p-x/z}{x/z}\right\rangle (x,z,Q^2) =
2 \left( 1-\frac{Q^2}{xs} \right)
\frac{z^2}{Q^2} \frac{\displaystyle \int_{-\infty}^{0} \d t\;  (-t)
(1-2\alpha(t)) f\left(
\frac{x}{z},t\right) }{\displaystyle \int_{-\infty}^{0} \d t\;
f\left(\frac{x}{z},t\right) } .
\end{equation}

\begin{figure}
\begin{center}
{}~ \epsfig{file=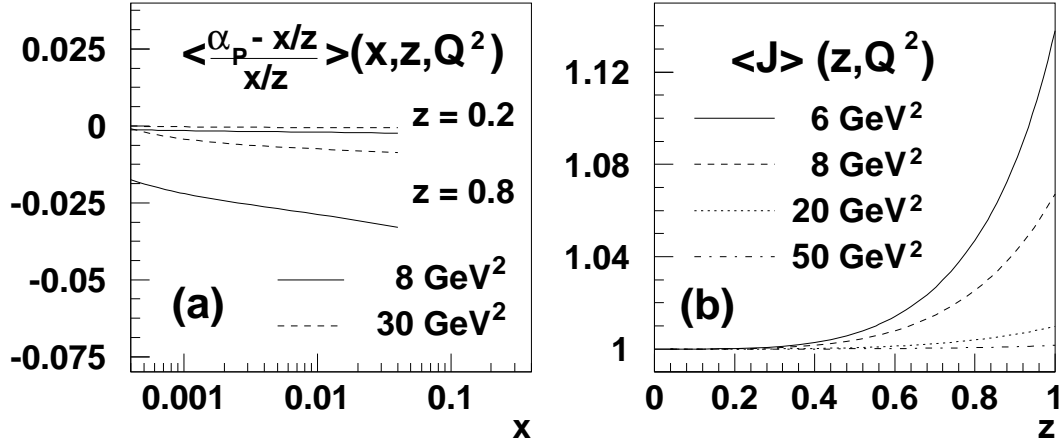,width=14cm}
\caption{Systematic deviations after averaging over $t$ and $\phi_{ep}$,
using the DL-parametrization for $f(\alpha_p,t)$ (\protect{\ref{fpom}}):
(a) systematic relative deviation between $\alpha_p$ and its
approximation $x/z$ as a function of $x$. The upper lines correspond
to $z=0.2$, the lower ones to $z=0.8$. $y_P$ and $y$ show the same
systematic deviations with the opposite sign; (b) magnitude of the
Jacobian factor defined in Eq.~(\protect{\ref{jacobian}}).}
\label{fig:sys}
\end{center}
\end{figure}

Figure~\ref{fig:sys}(a)  shows this systematic deviation for the
DL parametrization of $f(\alpha_p,t)$ (\ref{fpom}), which is
independent of the overall normalization of $f(\alpha_p,t)$ (the
pomeron flux). We see that
there is a small ($<$5\%) negative correction to the approximation
(\ref{h1appr}) for $\alpha_p$ and the same, positive, correction for
$y_P$. This effect can be understood
intuitively as follows. The form of $f(\alpha_p,t)$ favours low values
of $\alpha_p$ and therefore values of $90^{\circ}\leq \phi_{ep} \leq
270^{\circ}$, i.e. $\cos\phi_{ep} \leq 0$. In this region,
the pomeron moves towards the virtual
photon, thus increasing the virtual photon energy $y_P$ `seen' by the
pomeron.
Note that this effect decreases with increasing
$Q^2$ and so will vanish in the asymptotic scaling limit. Furthermore, the
deviation is proportional to the intercept $2\alpha(0)-1$,
which results in
corrections of up to 8\% for `hard' parametrizations of $f(\alpha_p,t)$,
as suggested from BFKL phenomenology \cite{bfklph}.

In order to examine the factorizability of the diffractive structure
function (\ref{f2ds}),
we return to Eq.~(\ref{diffX}) in its fully differential form:
\begin{eqnarray}
\label{fullX}
\frac{\d\sigma^{D\!S}}{\d\alpha_{p}\d t
\d x \d z \d Q^2} \; &  = &
\; \frac{4\pi\alpha^2}{xQ^4}\; \bigg\{
\frac{1+(1-y)^2}{2}\; F_2^{D\!S} (x,Q^2,\alpha_p,t) \nonumber \\
& & - \frac{y^2}{2} \;
F_L^{D\!S} (x,Q^2,\alpha_p,t) \bigg\} \int_{0}^{2\pi} \frac{\d
\phi_{ep}}{2\pi}\;  \delta (z-z(x,Q^2,\alpha_p,t)),
\end{eqnarray}
where we have made the replacement
\begin{equation}
R(x,Q^2,\alpha_p,t) = \frac{F_L (x,Q^2,\alpha_p,t)}{F_2
(x,Q^2,\alpha_p,t) - F_L (x,Q^2,\alpha_p,t)}.
\end{equation}
Assuming  that the factorization of the structure functions $F_2$ and
$F_L$ gives a valid approximation for the factorizability of the
cross section, this can be expressed as
\begin{eqnarray}
\frac{\d\sigma^{D\!S}}{\d\alpha_{p}\d t
\d x \d z \d Q^2} \; &  = &
\; \frac{4\pi\alpha^2}{xQ^4}\; \bigg\{
\frac{1+(1-y)^2}{2}\; F_2^{P} (z,Q^2,t) \nonumber \\
& & - \frac{y^2}{2} \;
F_L^{P} (z,Q^2,t) \bigg\} f(\alpha_p,t) \int_{0}^{2\pi} \frac{\d
\phi_{ep}}{2\pi} \delta (z-z(x,Q^2,\alpha_p,t)).
\end{eqnarray}
After a simple integration over $\phi$ and $x$, restricted to
the kinematically allowed values of the latter for fixed $\alpha_p$ and $z$,
we obtain finally an expression for the differential cross section
similar to (\ref{facexact}):
\begin{equation}
\label{facapprox}
\frac{\d\sigma^{D\!S}}{\d\alpha_{p}\d t
\d z \d Q^2} \;  =
\; \frac{4\pi\alpha^2}{zQ^4}\left(1+
\frac{4 t z^2}{Q^2}\right)^{-\frac{1}{2}}\; \left\{
1-y+\frac{y^2}{2[1+R^{P}(z,Q^2,t)]}\right\} \; F_2^{P}
(z,Q^2,t) f(\alpha_p,t).
\end{equation}
Assuming $F_2^P$ and $F_L^P$ to be independent of $t$, we can estimate
of the magnitude of the Jacobian factor,
\begin{equation}
\label{jacobian}
\langle J \rangle(z,Q^2,\alpha_p) \equiv
\left\langle\left(1+\frac{4 tz^2}{Q^2}\right)^{-\frac{1}{2}}  \right\rangle
 (z,Q^2,\alpha_p) =
\frac{\displaystyle \int_{-Q^2/4z^2}^{0} \d t\;  \left(1+
\frac{4tz^2}{Q^2}\right)^{-\frac{1}{2}} f(\alpha_p,t)}{\displaystyle
\int_{-\infty}^{0} \d t\;
f(\alpha_p,t)},
\end{equation}
which is shown in Fig.~\ref{fig:sys}(b) as a function of $z$ and
$Q^2$.\footnote{The $\alpha_p$ dependence turns out to be negligible.} We see
that this Jacobian factor differs by less than 5\% from unity for the whole
kinematical range experimentally accessible at HERA. Together with the
sytematic difference between
$y_P$ and $y$, which is of about the same order, we find that
the cross sections defined by (\ref{facexact})
and (\ref{facapprox}) agree within a  maximum deviation of 10\%, which
is attained
only  in the large-$z$ region. For values of $z<0.4$ the
agreement is already better than 5\%. Furthermore, both expressions become
equal in the scaling limit $Q^2\rightarrow \infty$. As the
experimental errors on the diffractive structure function are still
well above these corrections \cite{h1new},
 and uncertainties arising from the $R$-factor are
twice as big as these corrections, it seems appropriate at this time
to factorize
the diffractive structure function into a pomeron emission factor and
a deep inelastic structure function of the pomeron, Eq.~(\ref{f2ds}).
When, in the future,
the data improve and the full pomeron kinematics can be
reconstructed, it should be kept in mind that this factorization is
only an approximation to the factorization of the diffractive cross
section, Eq.~(\ref{facexact}).

A final  point concerns the measured intercept of the pomeron
trajectory. As a measurement of $t$ is not possible at present, only
an `average' coupling of the pomeron to the proton can be determined:
\begin{equation}
f(\alpha_p) = \int_{-\infty}^{0} \d t \; f(\alpha,t) \sim
\alpha_p^{-n(\mbox{eff})}.
\end{equation}
Using the DL-parametrization (\ref{fpom}) for $f(\alpha_p)$, one finds
$n(\mbox{eff})\simeq 1.09 \pm 0.02$. The error here
represents the spread in $n(\mbox{eff})$ values as $\alpha_p$
varies over the range $10^{-4} < \alpha_p < 10^{-2}$.
If $\alpha_p$ is approximated by $x/z$, the effective power increases
slightly  to
$n(\mbox{eff})\simeq 1.11 \pm 0.03$, which is a non-negligible shift.
Both these values are
significantly lower than the `naive'
approximation $n(\mbox{eff})\approx 1-2\alpha(0) =
1.17$, and therefore this effect should be taken into account in
comparing the measured intercept with  model predictions.

In summary, we have shown in this section that effects
arising from
an incomplete reconstruction of the pomeron kinematics at HERA
give systematic corrections of only a
few percent to $\alpha_p$, $y_P$ and the
measured intercept of the pomeron trajectory. Furthermore we have
demonstrated that the factorization of the diffractive structure function
gives a correct approximation of the factorization of the diffractive
cross section up to a relatively minor error, which vanishes in the
large $Q^2$ scaling limit.

\section{Final-state electron-proton correlations}
\label{sec:corr}

It should be clear from the above discussion that identification
of the scattered proton and measurement of its four momentum $p'$
will provide a crucial test of the pomeron picture. In principle, this would
allow a direct measurement of the parameters $\alpha_p$ and $t$
and hence of the pomeron emission factor $f$.
However in practice it will be difficult to make a precision
measurement of the proton energy, which would be needed to obtain
sufficient experimental resolution on $\alpha_p$ and hence a precise
determination of $t$. In  the short term, it therefore seems more
promising to test the $t$-dependence of $f$ by using
the angular correlation between the
transverse momenta of the outgoing electron and proton, Fig.~\ref{fig:one}.

As  discussed in the previous section, the $\alpha_p$ dependence of any
Regge-motivated $f(\alpha_p,t)$ favours low values of $\alpha_p$, and
therefore final state configurations in which the scattered electron
and proton are approximately back-to-back.
This effect will be enhanced with increasing
transverse momentum of the pomeron. Thus the distribution of
events in the relative azimuthal angle $\phi_{ep}$ is a  measure
of  the average scale of
$t$ involved in the process. The $\phi_{ep}$ dependence of the
diffractive cross section can be parametrized in  the
form of a distribution function:
\begin{equation}
\label{dndphi}
\frac{\d N}{\d \phi_{ep}} (x,z,Q^2,\phi_{ep}) = \frac{\displaystyle
\int_{t_{min}}^{0}
\d t \;f \left(\frac{x}{z}+2 x \sqrt{1-\frac{Q^2}{xs}}
\frac{\sqrt{-t}}{Q} \cos \phi_{ep} ,t \right)}{\displaystyle
\int_{-\infty}^{0} \d t
\;f\left(\frac{x}{z}, t\right)},
\end{equation}
where the lower limit on $t$ arises from the physical range of
the fractional proton momentum carried by the pomeron $0 \leq \alpha_p
\leq 1$:\footnote{This constraint is not to be confused with the more
restrictive experimental cuts on the quantity $x/z$, since
 $x$ and $z$ are fixed in this distribution.}
\begin{equation}
t_{min} = \left\{ \begin{array} {r@{\quad : \quad}l} \displaystyle
-\frac{Q^2}{4 z^2 (1-Q^2/xs) \cos^2 \phi_{ep}} &  90^{\circ}\leq
\phi_{ep} \leq 270^{\circ} \\ \displaystyle -\frac{Q^2}{4 (1-Q^2/xs)
\cos^2 \phi_{ep}} \left(\frac{1}{x} - \frac{1}{z} \right)^2 &
\phi_{ep} < 90^{\circ}, \phi_{ep} >  270^{\circ}. \end{array} \right.
\end{equation}
In practice, these bounds on $t$ have minimal impact on the
$\d N/\d \phi_{ep}$ distribution, since one expects $f$ to be
strongly suppressed for $|t|$ values larger than
 a `typical' hadronic scale of ${\cal O}(1\, \mbox{GeV}^2)$.

\begin{figure}
\begin{center}
{}~ \epsfig{file=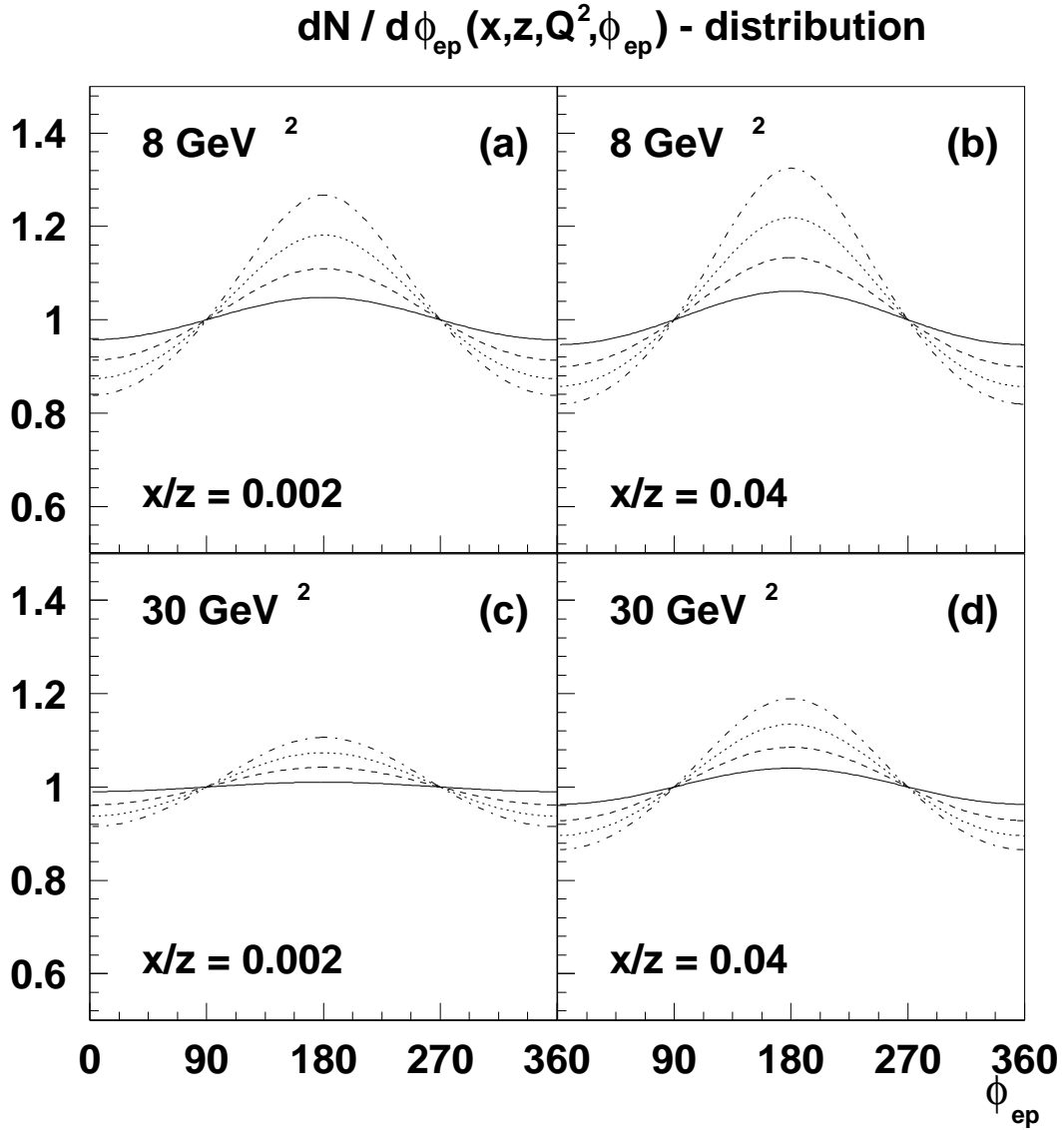,width=14cm}
\caption{$\d N/\d \phi_{ep} (x,z,Q^2,\phi_{ep})$ distribution for fixed values
of $x/z$ and $Q^2$, at
$z=0.2$ (solid lines),
$z=0.4$ (dashed lines),
$z=0.6$ (dotted lines),
 and $z=0.8$ (dot-dashed lines).}
\label{fig:dndphi}
\end{center}
\end{figure}

Figure~\ref{fig:dndphi} shows the predicted correlation between the
outgoing electron and the remnant proton as a function of $x$, $z$ and
$Q^2$. In fact it turns out that this function is almost independent of
the ratio $x/z$, the naive expectation for $\alpha_p$. As expected from
(\ref{dndphi}), the maximum asymmetry between the same-side and
opposite-side hemispheres is obtained for low values of $Q^2$ and high
values of $z$. Note that the  effect reaches a magnitude of up to 30\% for
realistic  HERA kinematics ($Q^2=8\,\mbox{GeV}^2, z=0.6$),  and hence should
be distinguishable from statistical fluctuations.

As we have discussed in detail in the previous section,
the discrepancy between
factorization at the level of the diffractive structure function and
the diffractive cross section is of order $-t/Q^2$, which is
subleading to the $\sqrt{-t}/Q$ dependence in (\ref{dndphi}).
It is therefore appropriate to use this angular distribution in
connection with the factorized structure function (\ref{f2ds}).
Assuming the structure functions $F_L^P$ and $F_2^P$ to be independent
of $t$, this yields the following result for the diffractive cross section:
\begin{equation}
\label{phisec}
\frac{\d\sigma^{D\!S}}{\d x \d z \d Q^2 \d \phi_{ep} } \;  =
\; \frac{4\pi\alpha^2}{z Q^4}\; \left\{
1-y+\frac{y^2}{2[1+R^{P}(z,Q^2)]}\right\} \; F_2^{P}
(z,Q^2) \frac{1}{2 \pi} \frac{\d N}{\d \phi_{ep}} (x,z,Q^2,\phi_{ep}).
\end{equation}
The  error implicit in  this expression due to the  neglect of the
Jacobian factor discussed in the previous section affects the
normalization of $\d N/\d \phi_{ep}$, and  leads to
\begin{equation}
\int_{0}^{2 \pi} \d \phi_{ep} \;
\frac{\d N}{\d \phi_{ep}} (x,z,Q^2,\phi_{ep}) > 2\pi.
\end{equation}
However  this deviation is less that 5\% for the kinematical range at
HERA, since it only reparametrizes the Jacobian factor
(\ref{jacobian}), which is small compared to the angular
asymmetry of up to 30\%.

Eq.~(\ref{phisec}) can be used to extract the $\d N/\d
\phi_{ep}$ distribution from the HERA data,  since it only requires
information on the coordinates of the remnant proton, and not on its
momentum. This distribution can provide a crucial test of the
applicability of  DL-like parametrizations of $f(\alpha_p,t)$.
Furthermore, any  $t$-dependence of $F_2^P$ would result in deviations from
the predicted $z$-dependence of $\d N/\d \phi_{ep}$.
In particular,  a significant
$t$-dependent contribution to $F_2^P$ would map the $z$-dependence of
$F_2^P$ onto the $z$-dependence of $\d N/\d \phi_{ep}$.

\section{Predictions for F$_2^{P}$ and F$_2^{D\!S}$}
\label{sec:qcd}

\subsection{Models for the partonic content of the pomeron}
\label{sec:partons}

The type and distribution of the parton constituents of the pomeron
has been a topic of some debate \cite{debate}.
On one hand, it seems natural to assume that
the pomeron is predominantly `gluonic' \cite{gluons}.
On the other hand, the pomeron must couple to quarks at some level.
In fact in Ref.~\cite{dldis} Donnachie and
Landshoff have presented a prediction
for the quark distribution in a pomeron
\begin{equation}
\label{zshape}
z q^P(z) \approx {\textstyle{1\over 3}} C \pi z (1-z) ,
\end{equation}
with    $C \approx 0.17$. This result is obtained from calculating
the box diagram for $\gamma^* P \to q \bar q$, in the same way
as the photon structure function is calculated in the parton
model from the box diagram for $\gamma^* \gamma \to q \bar q$.
A crucial difference for the above pomeron calculation is the softening
of the pomeron--quark vertex by a form factor which suppresses
large virtualities. This leads to the {\it scaling} behaviour (\ref{zshape})
in the $Q^2 \to \infty$ limit, in contrast to the asymptotic growth
$q^\gamma(x,Q^2) \sim a(x) \ln(Q^2 /  \Lambda^2)$ obtained for the  quark
distributions in the photon.

The absence of pointlike pomeron--quark couplings in the above model,
which gives rise
to asymptotic Bjorken scaling for the pomeron structure function,
suggests that the partonic content of the pomeron is on a similar footing
to that of any other hadron. In particular, we would expect
the parton distributions to satisfy a momentum sum rule, $f_q + f_g = 1$,
where $f_q$ ($f_g$) is the momentum fraction carried by quarks (gluons).
If we take the Donnachie-Landshoff form (\ref{zshape}) and assume
three light flavours of quarks and antiquarks, we find
$f_q = C\pi/3 = 0.18 \ll 1$, which
in turn suggests $g^P \gg q^P$. This is
the basis of the model of Ingelman and Schlein \cite{ing1}.
If the gluon and quark distributions in the pomeron are {\it both}
considered to be hard, as in Eq.~(\ref{zshape}) for example,
then good agreement is obtained with  the {\it shape}
of the UA8 jet rapidity distribution \cite{UA8}.
However, preliminary analysis of the UA8 data has shown that
the {\it normalization}
of the data appears to be a factor of about $2 - 5$ smaller
(depending on the assumed mixture of quarks and gluons) than the prediction
obtained from folding momentum-sum-rule constrained quark and gluon
distributions with pomeron emission factors such as that given by
Eqs.~(\ref{fpom},\ref{formfac}) \cite{SCHLEIN}.
There are two possible resolutions to this discrepancy.
\begin{itemize}
\item[{(i)}] It has been argued \cite{DLNPB,CTEQ,GOULIANOS} that
there is an ambiguity
in the definition of the pomeron emission factor $f(\alpha_p,t)$.
Since the pomeron is not a `real' particle, this factor cannot be interpreted
as a probability in the strict sense, and therefore its normalization
is process dependent. The normalization of the 'soft' $f$ differs
from the normalization of $f$ in any hard process. Even the
normalizations in different hard processes do not necessarily have to be
the same.
In the expression for the diffractive structure function
(\ref{f2ds}), one should therefore either replace the factor $f$
(obtained by a certain
procedure from `soft' hadronic cross sections) by $\calN f$,
or equivalently include an additional normalization factor $\calN$
in the relationship between the pomeron structure function
and the quark distributions, i.e. $F_2^P = \calN \sum_q e_q^2 q^P$.
We shall adopt the latter procedure below.

\item[{(ii)}] The pomeron flux factor is (approximately) universal, but the
pomeron has a pointlike coupling to quarks, which gives rise to a
direct contribution to the pomeron structure function.
The analogy here is with the photon structure function,
which grows asymptotically as $q^\gamma \sim \log Q^2$. A model
of pomeron structure of this type, i.e. with $q^P \sim \log Q^2$
has been presented in Ref.~\cite{KRAMER}.
\end{itemize}

We would like to propose two very  simple, physically motivated models
for the pomeron's parton structure based on each  of the above
assumptions. It will turn out that at present both are able to
give an adequate description of the diffractive structure function
measured by H1, but future more precise measurements, particularly of the $Q^2$
dependence, should be able to distinguish between them.

\subsubsection{Model~I}

We assume that at some bound-state scale, $Q_0^2=2\; \mbox{GeV}^2$
(corresponding roughly to the mass scale of the glueball candidate reported
in~\cite{WA91}), the pomeron is composed of valence gluons accompanied
by a small amount of valence quarks and antiquarks.\footnote{A similar
model, but with a softer gluon distribution, has been discussed
in \cite{CAPELLA}.}
As the pomeron carries the quantum numbers of the vacuum, its quark
and antiquark distributions have to be identical. Therefore, one only
has to consider two parton distributions in the pomeron, the quark
singlet  $\Sigma^P = \sum_i (q_i^P + \bar q_i^P)$
and the gluon. These are assumed to have the following,
valence-like shapes at $Q_0^2$:

\begin{equation}
\label{starting}
z \Sigma^P(z,Q_0^2) = f_q(Q_0^2)\; 6 z (1-z)  , \quad
z g^P(z,Q_0^2) = f_g(Q_0^2)\; 6  z(1-z) .
\end{equation}

For $Q^2 > Q_0^2$ additional quarks are generated dynamically, according
to the GLAP evolution equations of QCD \cite{glap},
and acquire a growing fraction of the pomeron's momentum.
In fact, leading-order perturbative QCD predicts that
the asymptotic momentum fractions are, regardless of the type of hadron,
\begin{equation}
f_q \rightarrow {3 n_f \over 16 + 3 n_f } , \quad
f_g \rightarrow { 16 \over 16 + 3 n_f }  .
\end{equation}
Our model is also motivated by the success of the dynamical parton model
for the proton structure functions~\cite{dynamical}, in which the proton
is a mixture of valence-like quarks and gluons~\cite{grv} at some low scale.

In the evolution of these parton distributions
we always define the quark
singlet to be the sum of only three light quark flavours ($u,d,s$);
contributions of heavy quarks to $F_2^P$, of which we will only
consider the dominant charm contribution, are incorporated by
projecting the massive
contribution from the $\gamma^{\star}g \rightarrow c \bar{c}$ fusion
process onto $F_2^P$. This treatment of heavy quark contributions to
deep inelastic structure functions has been shown~\cite{grvcharm} to be
more reliable than the construction of intrinsic heavy-quark
distributions in the hadron, which then evolve like massless partons
above a certain threshold. As argued in Ref.~\cite{grvcharm}, quark mass
effects clearly remain relevant even at energies above the HERA range,
which calls into question a massless resummation of these contributions.

For completeness, we will briefly outline the QCD treatment of the
light and heavy quark contributions
to the pomeron structure, although it is identical to the
procedure in Refs.~\cite{grvcharm,grvnew}.
The parton distributions at higher $Q^2$ are determined from the
leading-order\footnote{Although the full next-to-leading order
technology is available, we do not consider it to be appropriate in
this case. In the extraction of the diffactive structure function from
the experimental data~\protect{\cite{h1new}}, the diffractive
$R$-factor was set to zero, which can be only consistently accommodated
in a leading-order parton distribution model.} GLAP evolution equations
\begin{equation}
 \label{glap}
\frac{\partial}{\partial\ln Q^2}
\left(\begin{array}{c} q_i^P(z,Q^2) \\ g^P(z,Q^2) \end{array}\right)
 = \frac{\alpha_s (Q^2)}{2\pi}
\int^1_z {d\xi \over \xi} \;
\left(\begin{array}{cc}P_{qq}\left({z\over \xi}\right)
& P_{qg}\left({z\over \xi}\right)  \\
                       P_{gq}\left({z\over \xi}\right)
& P_{gg}\left({z\over \xi}\right)
\end{array}\right)
\left(\begin{array}{c}q_i^P(\xi,Q^2) \\ g^P(\xi,Q^2)\end{array}\right),
\end{equation}
in which we keep the number of massless flavours fixed at $n_f=3$ in the
splitting functions $P_{ij}$, while the number of active flavours in
the running of $\alpha_s(Q^2)$ is determined by the $Q^2$ scale.
This procedure results in continuous parton distributions and
couplings at each flavour threshold, while $\Lambda_{LO}^{QCD}$ is
matched at each threshold. We use
$\Lambda_{\rm LO}^{\rm QCD}(n_f=4)=200\;\mbox{MeV}$.

Assuming that SU(3) flavour symmetry is already
established at $Q_0^2$, the contribution of the light quarks flavours
to $F_2^P$ is just the singlet distribution times a charge factor:
\begin{equation}
F_2^{P(u,d,s)}(z,Q^2) = \frac{2}{9} z \Sigma(z,Q^2).
\label{lightf2}
\end{equation}
The massive charm contribution arising from photon-gluon fusion
takes the form
\begin{equation}
\label{f2charm}
F_2^{P(c)}(z,Q^2,m_c^2) = 2 z q_c^2 \frac{\alpha_s(\mu_c^2)}{2 \pi}
\int_{az}^1 { \d y \over y} \; C\left( \frac{z}{y}, {m_c^2\over Q^2} \right)
\; g^P(y,\mu_c^2) ,
\end{equation}
with the kinematical bound $a = 1 + 4 m_c^2/Q^2$ and the LO coefficient
function
\begin{eqnarray}
C(\zeta, r) & = & \frac{1}{2}[\zeta^2 + (1-\zeta)^2 + 4\zeta(1-3\zeta) r -
8\zeta^2 r^2] \ln{1+\beta\over 1 -\beta}  \nonumber \\
& & + \frac{\beta}{2} [-1 + 8\zeta(1-\zeta)-4\zeta(1-\zeta)r] ,
\label{coefch}
\end{eqnarray}
where
\begin{equation}
\beta^2 = 1 - {4 r \zeta \over 1-\zeta} .
\end{equation}
It has been shown in~\cite{grvcharm} that a mass factorization scale of
$\mu_c^2=4m_c^2$ for the gluon distribution in the above expression
is the most appropriate choice with regard to the perturbative
stability of the expression. We will use
$m_c=1.5\;\mbox{GeV}$ in our numerical evaluations presented below.
The complete prediction for $F_2^P$ is therefore
\begin{equation}
\label{withcharm}
F_2^P(z,Q^2)  = {\textstyle{2 \over 9}} z \Sigma^P(z,Q^2)
+ F_2^{P(c)}(z,Q^2,m_c^2) .
\end{equation}
The above treatment of the charm contribution takes proper account
of the threshold behaviour which, as we will see in
Section~\ref{sec:data}, makes a significant contribution to the $Q^2$
dependence of the structure function.

Finally, to take into account the ambiguity in the normalization
of the pomeron flux factor discussed above, we multiply the structure
function (\ref{withcharm}) by an overall normalization factor $\calN$,
\begin{equation}
\label{rescaled}
F_2^P(z,Q^2)  \longrightarrow \calN F_2^P(z,Q^2)  .
\end{equation}
We shall see that $\calN \approx 2$ gives a good representation
of the H1 data.  Note that this factor cannot be directly compared to the
`discrepancy factor' obtained in the UA8 jet analysis
\cite{SCHLEIN}, since hard diffractive hadron-hadron processes do not general
possess the factorization properties which would lead to universality
 of the pomeron emission factor \cite{CTEQ,CFS,soper}.

\subsubsection{Model~II}
 In the second approach discussed above, the pomeron structure function
 has two distinct contributions, a `resolved' component which consists
 of hadron-like quark and gluon constituents, as in Model~I, and a `direct'
 component coming from an assumed pointlike coupling of the pomeron
 to the quarks. This is in exact analogy to the photon structure function,
 and there is consequently no overall momentum sum rule. The
Donnachie-Landshoff
 flux factor is used without modification.

The
resolved contribution is described in terms of parton distributions
evolving from valence-like distributions for light quarks and gluons
at a bound state scale of $Q_0^2=\; 2 \;\mbox{GeV}^2$:
\begin{equation}
z \Sigma^P(z,Q_0^2) = f_q(Q_0^2)\; 6 z (1-z)  , \quad
z g^P(z,Q_0^2) = f_g(Q_0^2)\; 6  z(1-z) .
\end{equation}
The contribution of the light resolved quarks to  $F_2^P$ is
then:
\begin{equation}
F_2^{P(u,d,s),{\rm res.}}(z,Q^2) = \frac{2}{9} z \Sigma^{{\rm
res.}}(z,Q^2).
\label{lightf22}
\end{equation}
The resolved charm contribution arising from the fusion of a resolved gluon
with the photon takes the same form as in Model~I (\ref{f2charm}):
\begin{equation}
\label{reschm}
F_2^{P(c),{\rm res.}}(z,Q^2,m_c^2) = F_2^{P(c)}(z,Q^2,m_c^2)
\end{equation}

The direct coupling of the pomeron to quarks with an unknown coupling
$c$ gives rise to a pointlike contribution to $F_2^P$ from
pomeron-photon fusion.\footnote{The treatment
presented here is similar to the discussion in Ref.~\cite{KRAMER}.}
 Here we take only the nonresummed contribution from
the corresponding parton-level subprocesses, as the resummed distributions
tend to overestimate the magnitude of the distributions in the
high-$z$ region.
The light-quark contribution to the structure function is then
\begin{equation}
\label{ldf2}
F_2^{P(u,d,s),{\rm dir.}}(z,Q^2) = 2 \sum_q \;e_q^2 \;\frac{3}{8\pi^2} \;c^2\;
z \left[ z^2 + (1-z)^2 \right] \ln \frac{Q^2(1-z)}{m_q^2 z} =
\frac{2}{9} z \Sigma^{{\rm dir.}}(z,Q^2).
\end{equation}
Here we assume the mass regulator for all light quark masses to be
$m_q =0.3\;\mbox{GeV}$. The full contribution from charm quarks above
their production threshold
\begin{equation}
\label{thre}
Q^2 > 4 m_c^2 \frac{z}{1-z}
\end{equation}
takes the form
\begin{equation}
\label{dirchm}
F_2^{P(c),{\rm dir.}}(z,Q^2,m_c^2) = 2 \;q_c^2  \;\frac{3}{4\pi^2} \;c^2\;
 C\left(z,\frac{Q^2}{m_c^2}\right).
\end{equation}

The structure function of the pomeron is therefore
\begin{eqnarray}
F_2^P (z,Q^2) & = & F_2^{P(u,d,s),{\rm res.}}(z,Q^2) + F_2^{P(c),{\rm
res.}}(z,Q^2,m_c^2) \nonumber \\
\label{f2m2}
& & + F_2^{P(u,d,s),{\rm dir.}}(z,Q^2)+ F_2^{P(c),{\rm dir.}}(z,Q^2,m_c^2).
\end{eqnarray}

We shall see below that both models can be adjusted to give
satisfactory agreement with the recent H1 data, but give significantly
different predictions for the $Q^2$-dependence of $F_2^P$ and for the
charm content of diffractive scattering events.

\subsection{Q$^2$ evolution  of F$_2^{D\!S}$}

\label{sec:evo}
The assumption that $F_2^{D\!S}$ is factorizable into an emission and a
DIS part (\ref{f2ds}) implies that the $Q^2$ dependence
of $F_2^{D\!S}$ arises entirely from $F_2^{P}$. Furthermore, if the parton
interpretation of $F_2^{P}$ is valid, then this $Q^2$ dependence should
be given by the standard GLAP evolution equations (\ref{glap})
of perturbative QCD. The observation of this
$Q^2$ dependence is an important test of the whole approach, in
particular of the factorizability of diffractive scattering cross
sections.

The $Q^2$ evolution of the resolved contribution to
 $F_2^{P}$ is given directly by the GLAP
equations (\ref{glap}).
For $F_2^{D\!S}$ we must fold the
results with the pomeron flux factor $f$.
In particular we can define `diffractive' parton distributions
in the proton by
\begin{equation}
\label{dsparton}
\left(\begin{array}{c} q^{D\!S}(x,Q^2,t)
                    \\ g^{D\!S}(x,Q^2,t) \end{array}\right)
 = \int_0^1 \d z \d \alpha_p\;
\left(\begin{array}{c} q^P(z,Q^2,t)
                      \\ g^P(z,Q^2,t) \end{array}\right)
    f(\alpha_P,t) \delta(z-x/\alpha_p),
\end{equation}
where we have used (\ref{master}), dropping the small corrections due to
finite $t$ and $M^2$  effects.
Taking $\partial/\partial\ln Q^2$ of both sides, and using the fact
that the resolved pomeron parton distributions satisfy the GLAP equation,
Eq.~(\ref{glap}), gives
\begin{eqnarray}
\frac{\partial}{\partial\ln Q^2}
\left(\begin{array}{c} q^{D\!S}(x,Q^2,t)
                    \\ g^{D\!S}(x,Q^2,t) \end{array}\right)
 &=& \frac{\alpha_s (Q^2)}{2\pi} \;\int_0^1 \d y' \d \eta \d z \d \alpha_p\;
\; \left[ {\cal P}(\eta) \right]\;
\left(\begin{array}{c} q^P(y',Q^2,t)
                      \\ g^P(y',Q^2,t) \end{array}\right) \nonumber \\
  & & \times  f(\alpha_P,t)\; \delta(z-x/\alpha_p) \; \delta(z-\eta y'),
\label{dspartonevo}
\end{eqnarray}
where $\left[ {\cal P}(\eta) \right] $ is the $2 \times 2$ matrix of
splitting functions.
Introducing $ 1 = \int \d y \delta(y-y'\alpha_p)$ and integrating over
$y'$ and $z$ gives
\begin{equation}
\label{dspartonevox}
\frac{\partial}{\partial\ln Q^2}
\left(\begin{array}{c} q^{D\!S}(x,Q^2,t)
                    \\ g^{D\!S}(x,Q^2,t) \end{array}\right)
 = \frac{\alpha_s (Q^2)}{2\pi}\;  \int_x^1 \frac{\d y}{y} \;
 \left[ {\cal P}(x/y) \right]\;
\left(\begin{array}{c} q^{D\!S}(y,Q^2,t)
                    \\ g^{D\!S}(y,Q^2,t) \end{array}\right)
\end{equation}
which is the usual GLAP equation, but now for the resolved
diffractive parton distributions.

If the pomeron has a direct coupling to quarks as discussed in
Model~II, this will result in a direct contibution to the diffractive
structure function
\begin{equation}
\label{f2dds}
F_2^{D\!S,{\rm dir.}}(x,Q^2,t) = \int_0^1 \d z \d \alpha_p F_2^{P,{\rm
dir.}}(z,Q^2) \; f(\alpha_P,t) \delta(z-x/\alpha_p).
\end{equation}
In the scaling limit $Q^2\rightarrow \infty$, this term will dominate
the $Q^2$-dependence of $F_2^{D\!S}$
\begin{equation}
\label{evodir}
\frac{\partial}{\partial\ln Q^2} F_2^{D\!S,{\rm dir.}}(x,Q^2,t)=
 2 \sum_q \;e_q^2 \;\frac{3}{8\pi^2} \;c^2\;
\int_x^1 \d \alpha_p \;
\frac{x}{\alpha_p} \left[ \left(\frac{x}{\alpha_p}\right)^2 +
\left(1-\frac{x}{\alpha_p}\right)^2 \right]  f(\alpha_p,t),
\end{equation}
where the sum runs over all flavours above their production threshold
(\ref{thre}).

Therefore, one
should find that the $Q^2$ dependence of both $F_2^{D\!S}(x,Q^2,t)$ {\it
and}  $F_2^P(z,Q^2,t)$ is consistent with perturbative QCD
while the corresponding parton distributions are related by
Eq.~(\ref{dsparton}). A possible direct contribution to $F_2^P$ will
dominate the $Q^2$-dependence of the diffractive structure function
and should therefore be identifiable as the data improve.

It is worth stressing that the $Q^2$ dependences of
the proton structure function $F_2$ and
$F_2^{D\!S}$ {\it at the same Bjorken $x$ value}
are completely unrelated.
In particular,  $F_2$ rises rapidly with
increasing $Q^2$ at small $x$ as more and more slowly-moving
partons are generated by  the branching process. This rise is observed
\cite{HERA} to be proportional to $\ln Q^2 $ at fixed $x$, which is
consistent with recent parametrizations \cite{grvnew,mrsg,cteq3} of the
parton densities in the proton.
In contrast, the resolved quarks in the pomeron
are sampled at $z$ values much larger than $x$, where the distributions
evolve more slowly.

For a pomeron without a direct coupling to quarks (Model I),
one should therefore find that the fraction of diffractive events at
fixed $x$ is decreasing approximately like $1/\ln Q^2 $. A direct
coupling (Model II) will result in a rise of $F_2^{D\!S}$
proportional to $\ln Q^2 $ at fixed $x$. The intercept of this rise
is unrelatedto the rise of the proton structure function and could
in principle be
used to determine the coupling strength of the pomeron to quarks from
Eq.~(\ref{evodir}).

\section{Comparison with data}
\label{sec:data}
The H1 collaboration has recently measured \cite{h1new}
the diffractive  structure function $F_2^{D(3)}$
\begin{equation}
\label{diffh1}
\frac{\d\sigma^{D\!S}}{
\d x \d Q^2\d x_{P} } \;  =
\; \frac{4\pi\alpha^2}{xQ^4}\; \frac{1+(1-y)^2}{2}\; F_2^{D(3)}
(\beta,Q^2,x_P)
\end{equation}
as a function of the three kinematic variables, $\{\beta,Q^2,
x_P\}$ where
\begin{eqnarray}
\beta &=& { Q^2 \over M_X^2 + Q^2}\  \approx \ z \nonumber \\
x_P & = & {x\over \beta}\  \approx \ \alpha_P
\end{eqnarray}
with the approximations becoming exact when $t = M^2 = 0$.
The variables $\alpha_P$, $t$ and $\phi_{ep}$ are not measured
directly. By implication $0 \leq \phi_{ep} \leq 2 \pi$, and it is
estimated that $|t| \lapprox 7$~GeV$^2$ \cite{h1new}.

By integrating Eq.~(\ref{diffX}) we obtain our prediction for the measured
diffractive structure function
\begin{equation}
\label{prediction}
F_2^{D(3)}(\beta,Q^2,x_P) =  \int_0^1 \d \alpha_p
\int_{t_{\rm min}}^{t_{\rm max}}
\d t \int_0^{2\pi} {\d\phi_{ep}\over 2\pi}\;
\delta\left(x_P - {x\over z} - {xt\over Q^2} \right)\;
f(\alpha_p,t)\; F_2^P(z,Q^2,t) ,
\end{equation}
with  $z$ given in terms of the other variables by (\ref{master}).
Ignoring the $t$ dependence everywhere except in $f$, and setting
the proton mass to zero, we obtain the simple factorizing
approximation
\begin{equation}
F_2^{D(3)}(\beta,Q^2,x_P) \approx
\left[ \int_{t_{\rm min}}^{t_{\rm max}} \d t \;
f(x_P,t)\right] \; F_2^P(\beta,Q^2) ,
\end{equation}
which implies that the dependence of the structure function
on $x_P$ should be universal, i.e. independent of $\beta$ and $Q^2$.
Furthermore, if we substitute for $f$ using Eq.~(\ref{fpom}) we find
\begin{equation}
\label{scaling}
F_2^{D(3)}(\beta,Q^2,x_P) \approx
K\; x_P^{-n} \;  F_2^P(\beta,Q^2) .
\end{equation}
Precisely this behaviour has recently been observed
 by the H1 collaboration \cite{h1new}. In fact their measured
`universal' power $n$ of $x_P$ is $ n =  1.19 \pm 0.06(stat.)
\pm 0.07(sys.)$, which is in excellent agreement with our
 prediction of $1.11\pm 0.03$ (see Section~2)
  based on a correct
 treatment of kinematics and using the pomeron
emission factor of  Donnachie and Landshoff \cite{dldis}.

The H1 collaboration  have also attempted to measure the pomeron structure
function directly, by defining an $x_P$-integrated diffractive
structure function
\begin{equation}
\label{h1f2p}
\tilde{F}_2^{D}(\beta,Q^2) = \int_{0.0003}^{0.05} \d x_P \;
F_2^{D(3)}(\beta,Q^2,x_P),
\end{equation}
where the range of integration is chosen to span the entire $x_P$
measurement range\footnote{Very recently a similar measurement has
also been made by the ZEUS collaboration \cite{newzeus}. The
$\beta,Q^2$ dependence is consistent with the H1 data. The
normalization is different, however, because of the different $x_P$
range ($0.00063 < x_P < 0.01$), which results in $A$({\rm
ZEUS})$\approx 0.8$.}.
According to the simple factorization hypothesis, the diffractive
structure function is proportional to the pomeron structure function:
\begin{equation}
\label{relation}
\tilde{F}_2^{D}(\beta,Q^2) \approx A \;
F_2^{P}(\beta,Q^2),
\end{equation}
with
\begin{equation}
\label{Anumber}
 A =  \int_{0.0003}^{0.05} {\rm d} \alpha_p
\int_{t_{\rm min}}^{t_{\rm max}} {\rm d} t \;
f(\alpha_p,t) \approx 1.5.
\end{equation}
The numerical value in (\ref{Anumber}) corresponds
to the DL form
for $f$. In what follows we will use Eq.~(\ref{relation}) with $A = 1.5$
to convert the measured structure function \cite{h1new}
into the pomeron structure function.

In Ref.~\cite{h1new}, data on $\tilde{F}_2^{D}(\beta,Q^2)$ are presented
in four $Q^2$ bins, $Q^2 = 8.5,\; 12,\; 25,\; 50$~GeV$^2$.
In the first of these, the charm contribution should be relatively small,
and hence $\tilde{F}_2^{D}(\beta,Q^2)$ can be directly
compared with the predictions of our simple models for the light
quark distributions, which allows us to tune the parameters of both
models.

\subsection{Model~I}

\begin{figure}
\begin{center}
{}~ \epsfig{file=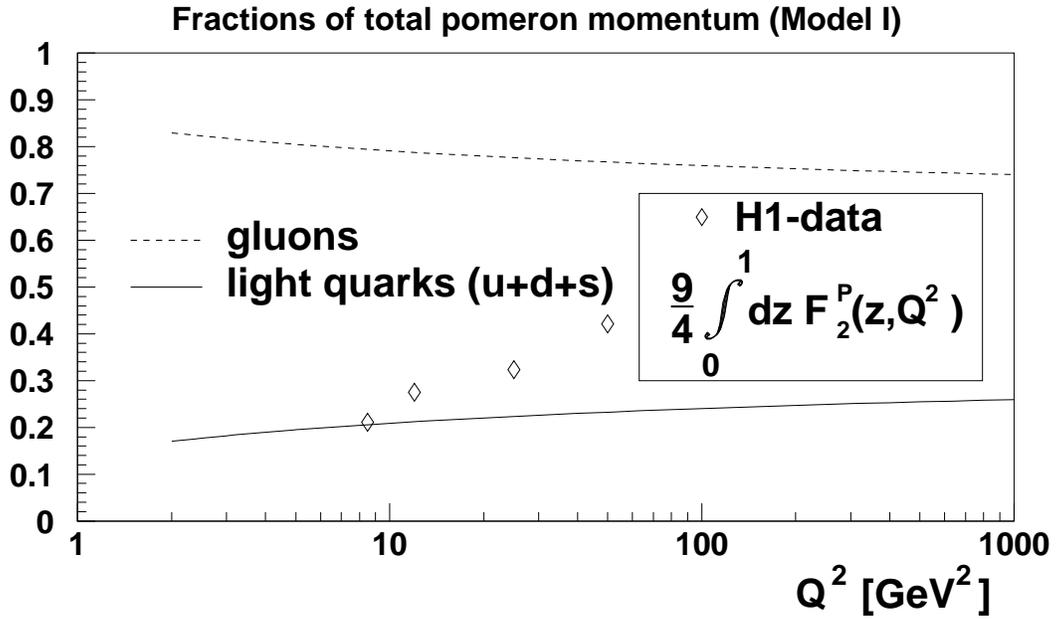,width=14cm}
\caption{Fractions of total pomeron momentum carried by light quarks
and gluons as predicted by leading-order GLAP evolution for three
light flavours in Model~I (see text).
The H1 datapoints shown in comparison are the values for the momentum
fraction carried by the sum of all light quarks under the naive assumption
of a negligible direct charm contribution to $F_2^P$. A
normalization of ${\cal N}=2$ for the pomeron
emission factor of Donnachie and Landshoff  is assumed.}
\label{fig:pfrac}
\end{center}
\end{figure}

As the pomeron fulfils a momentum sum rule in this model, the first
moment of $\tilde{F}_2^{D}$, which is related to the
the momentum fraction carried by quarks in the pomeron, contains the
information needed to adjust the main parameters. Neglecting the charm
contribution to $F_2^P$, this relation reads:
\begin{equation}
\label{h1momfrac}
 A^{-1} \; \int_0^1 \d\beta\; \tilde{F}_2^{D}(\beta,Q^2)
 \approx \int_0^1 \d\beta\; F_2^{P}(\beta,Q^2)
 \approx \frac{2}{9} {\cal N} \int_0^1 \d z\; z\Sigma^{P}(z,Q^2)
 =  \frac{2}{9} {\cal N} f_q(Q^2).
\end{equation}
The parameters  ${\cal N}$ and $f_q(Q^2_0)$
are strongly correlated --- their product is essentially
determined by the first moment
of $\tilde{F}_2^{D}$ in the lowest $Q^2$ bin.
Allowing for only a small charm contribution
in this bin, we find the best agreement with the data for
${\cal N}=2$ and the
following momentum fractions of quarks and gluons at $Q_0^2$:
\begin{equation}
f_q(Q_0^2) = 0.17, \quad f_g(Q_0^2) =0.83.
\end{equation}
Fig.~\ref{fig:pfrac} shows the values of $f_q$ extracted from the H1 data
\cite{h1new} in this way\footnote{The $\beta$-integrated structure
function in (\ref{h1momfrac}) is estimated by assuming that the structure
function is independent of $\beta$ at each $Q^2$ value.
This is a very crude
procedure, and we have no way of estimating the errors on the integral
obtained by this method. Our comparison is therefore only
semi-quantitative at best.}
at the four $Q^2$ values. Note that in the measured
$Q^2$ range, the momentum fractions are predicted to
vary only slightly with $Q^2$.
The apparent rise in the data has a simple interpretation as the onset
of the charm contribution, as predicted by (\ref{withcharm}).

\begin{figure}
\begin{center}
{}~\epsfig{file=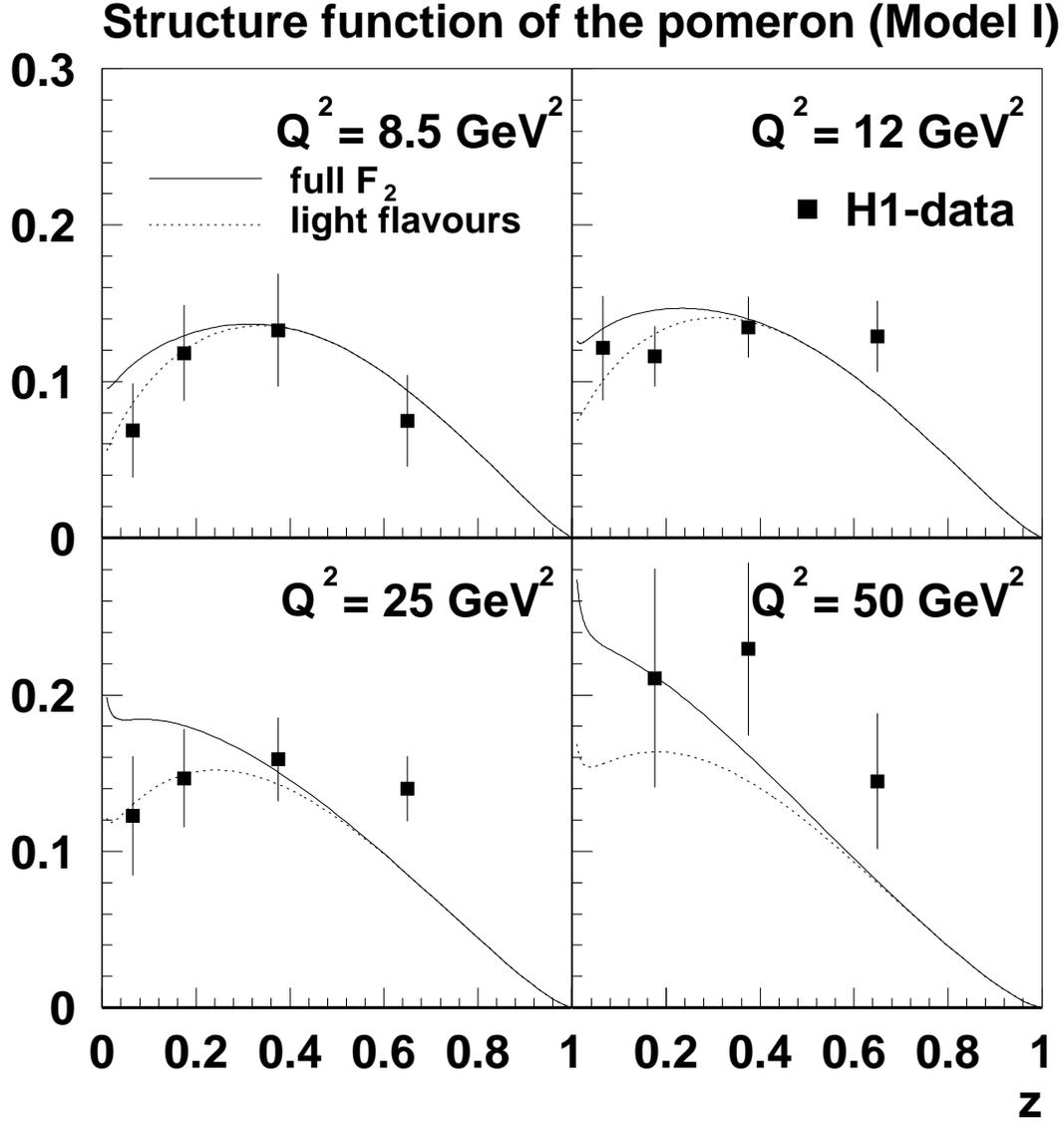,width=14cm}
\caption{The deep inelastic structure function of the pomeron $F_2^P(z,Q^2)$
constructed from the parton distributions in Model~I (described in the text)
of the three light-quark
flavours (dotted line) and with an additional charm
contribution from the photon-gluon fusion process (solid line).
The H1 data are obtained from values for the diffractive
structure function in terms of these variables~\protect{\cite{h1new}},
divided by a pomeron emission factor of 1.5 (derived from the model
of Donnachie and Landshoff). The theoretical predictions are scaled by
a  factor ${\cal N}=2$, as discussed in the text.}
\label{fig:f2p1}
\end{center}
\end{figure}

In Fig.~\ref{fig:f2p1}
the predictions of this model for the pomeron
structure function are compared with the data, as defined by
(\ref{relation}). The solid curves show the full prediction including
the charm contribution, and the dotted curves are the
contributions from the three light quarks only. We note that  for this model,
\begin{itemize}
\item[{(i)}] an accurate description of the data without a direct
pomeron coupling seems to be possible if the pomeron flux is adjusted;
\item[{(ii)}] the variation of the dotted curves with $Q^2$ shows that
the scaling violations predicted by the QCD evolution equations are rather
weak in this kinematic range;
\item[{(iii)}] the charm contribution grows rapidly above threshold
(in fact, this growth is
evidently responsible for the bulk of the predicted $Q^2$ dependence),
and constitutes a significant fraction of the
structure function at high $Q^2$
and low $z$;
\item[{(iv)}] as $Q^2$ is increased to higher values,
the pomeron structure function is
expected to rise rapidly at low $z$ and to decrease slowly at high $z$.
\end{itemize}

\begin{figure}
\begin{center}
{}~ \epsfig{file=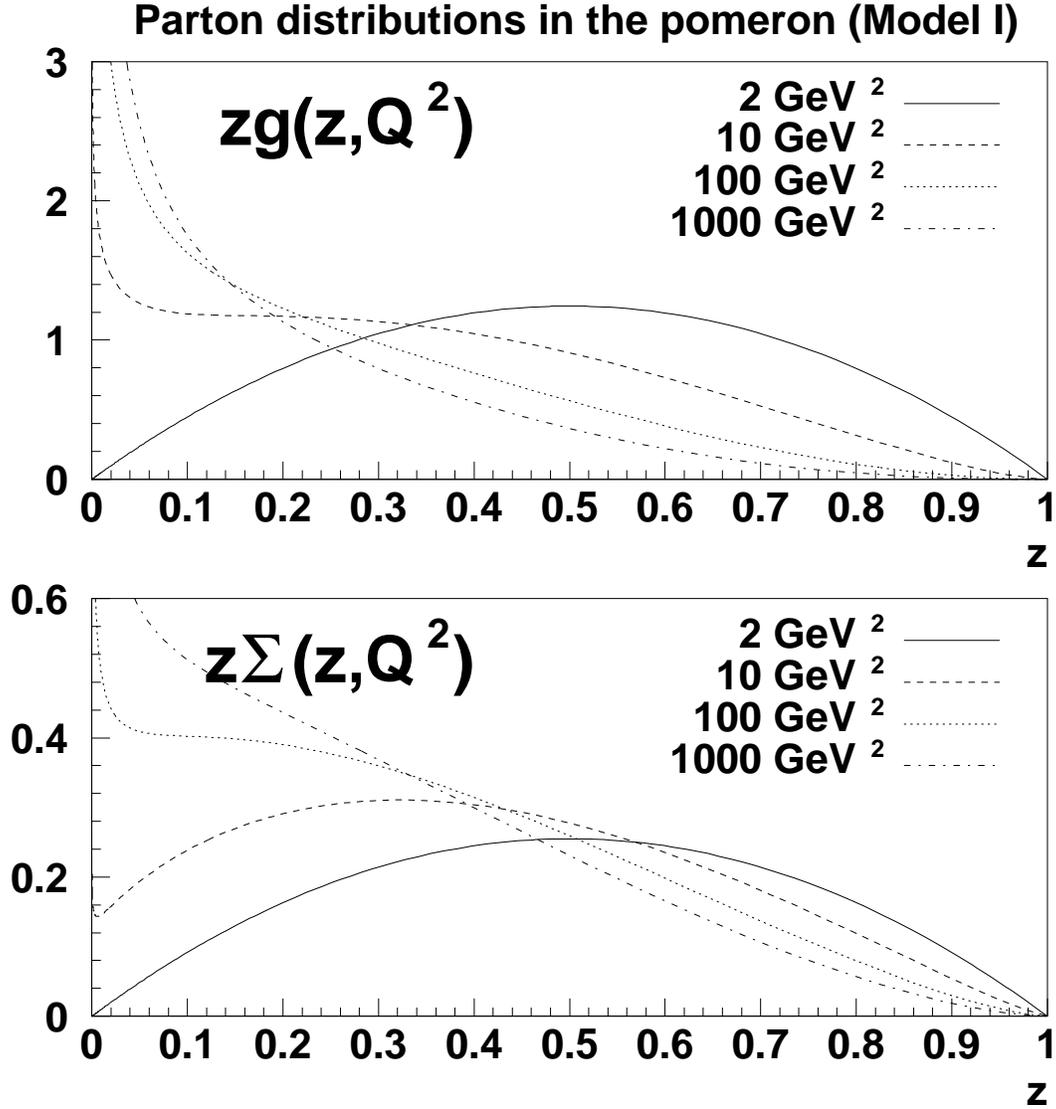,width=14cm}
\caption{Parton distributions in the pomeron, assuming a valence-like
structure at $Q_0^2=2\;\mbox{GeV}^2$ and no pointlike coupling of the
pomeron to quarks (Model~I). The relative normalizations are
chosen such that gluons carry 83\% and light quarks carry 17\% of the pomeron's
momentum at $Q_0^2$.}
\label{fig:partons1}
\end{center}
\end{figure}

Finally, in Fig.~\ref{fig:partons1} we show the gluon and singlet
(light) quark distributions in the pomeron,
as predicted in this model.\footnote{The {\tt FORTRAN} code
for the distributions and structure functions in both models is available by
electronic mail from T.K.Gehrmann@durham.ac.uk} Since
we are assuming exact SU(3) flavour symmetry, the individual quark
or antiquark distributions are simply $q^P = {\textstyle{\frac{1}{6}}}
\Sigma^P$.  Note that as $Q^2$ increases, both
the quark and gluon distributions evolve slowly to small $z$,
as expected. The emergence of a small-$z$ `sea'
of $q \bar q$ pairs can be seen at high $Q^2$.

\subsection{Model~II}

In order to adjust the free parameters of this model (the ratio of
resolved quarks and resolved gluons at $Q_0^2$ and the strength of the
direct coupling $c$), we use the fact that the direct contribution
(\ref{ldf2}) is concentrated in the high-$z$ region. Therefore, the
$z=0.065$ data points can be used to adjust the resolved contribution
to the pomeron structure function. Best agreement is found for
\begin{equation}
f_q(Q_0^2) = 0.3, \quad f_g(Q_0^2) =0.7.
\end{equation}
Having fixed the resolved contribution, the quark-pomeron coupling
constant is determined from the data to be $c=1$. The parameters of
this model are almost identical to the parameters of a similar model
described in Ref.~\cite{KRAMER}, obtained from an analysis of the rapidity and
transverse energy distributions in diffractive photoproduction.

\begin{figure}
\begin{center}
{}~\epsfig{file=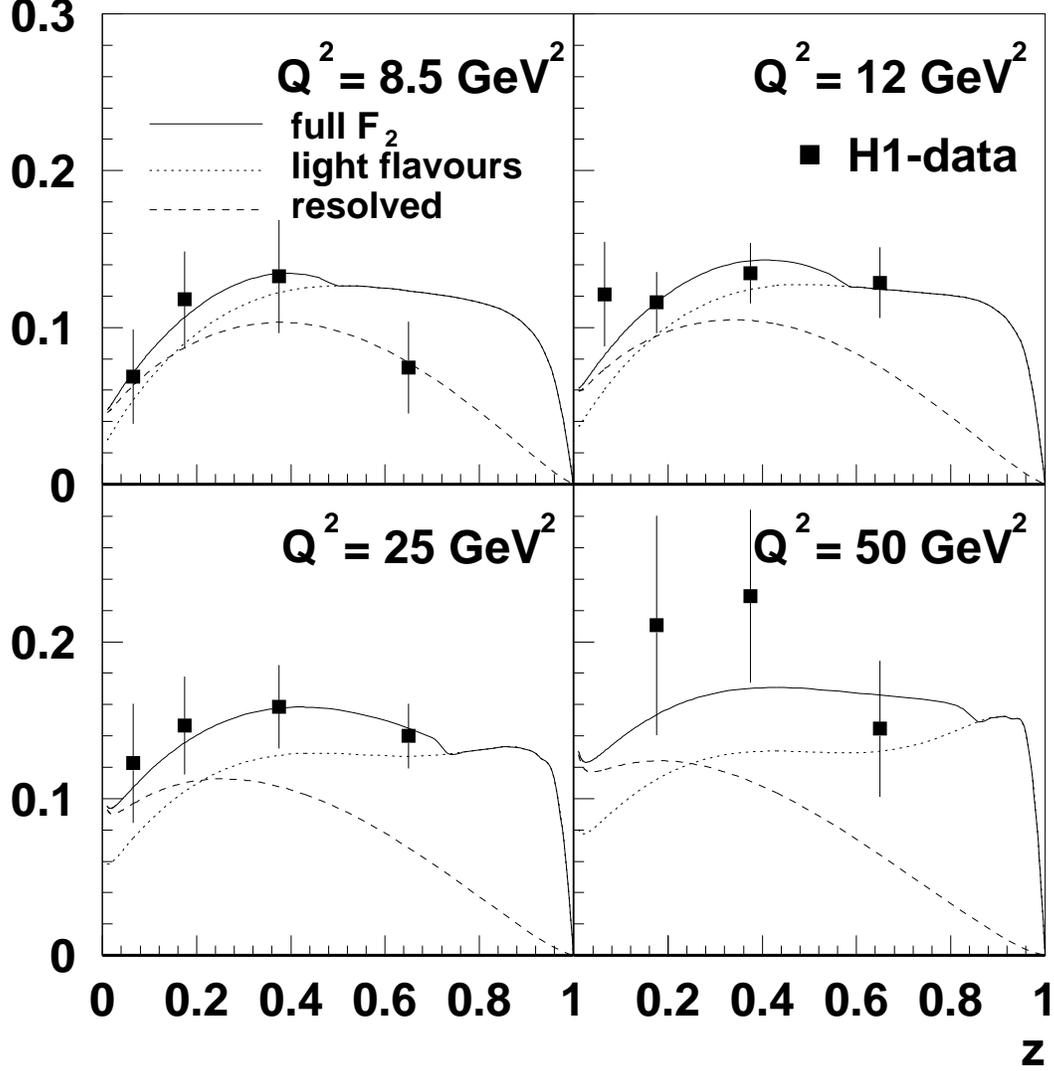,width=14cm}
\caption{The deep inelastic structure function of the pomeron $F_2^P(z,Q^2)$
constructed from the parton distributions
in Model~II (described in the text)
for three light-quark
flavours (dotted line) and with an additional charm
contribution from the photon--gluon and photon--pomeron fusion
processes (solid line). The dashed line repesents the resolved
contribution to $F_2^P$ for all four quark flavours.
The H1 data are obtained from values for the diffractive
structure function in terms of these variables~\protect{\cite{h1new}},
divided by a pomeron emission factor of 1.5 (derived from the model
of Donnachie and Landshoff). This emission factor is taken literally
to be the pomeron flux factor.}
\label{fig:f2p2}
\end{center}
\end{figure}

Figure~\ref{fig:f2p2} shows the predictions of this model for the
pomeron structure function compared with the H1 data. The solid line
represents the full $F_2^P$, Eq.~(\ref{f2m2}), the dotted line is the
sum of the resolved (\ref{lightf22}) and direct (\ref{ldf2}) light flavour
contributions and the dashed line gives the resolved contribution
summed over light (\ref{lightf22}) and charm (\ref{reschm}) quarks. We
note that for this model,
\begin{itemize}
\item[{(i)}]  a good description of the H1 data is obtained --- in particular,
 the agreement in the $Q^2=25 \ \mbox{GeV}^2$ bin is better than for
Model~I;
\item[{(ii)}] the pomeron structure function grows with $Q^2$ over the
whole range in $z$, due to  the dynamical generation of a
$q\bar{q}$ sea at low $z$ and the $P\rightarrow q\bar{q}$ splitting
at high $z$;
\item[{(iii)}] the charm contribution grows rapidly above threshold;
in contrast to Model~I, where the bulk of the charm events is
concentrated at low $z$, we find a sizeable charm contribution in the
high-$z$ region.
\end{itemize}

The different charm distribution in both models can be easily
understood from the analytic forms of $F_2^{P(c)}$
(Eqs.~(\ref{f2charm}), (\ref{reschm}) and (\ref{dirchm})): the coefficient
function for the fusion of a virtual photon with a spin-1 boson to
produce a
massive quark-antiquark pair (\ref{coefch}) is the same (up to colour and
normalization factors) in all three contributions. In Model~I and for
the resolved contribution to Model~II, this coefficient
function is convoluted over the distribution of gluons in the pomeron.
As this distribution  vanishes for $z\rightarrow 1$, one will only
find a small charm contribution directly below the threshold. The charm
production from pomeron--photon fusion in Model~II involves an
(implicit) convolution of the above coefficient function with the
distribution of `pomerons in the pomeron', which is $\delta(1-z)$ at
lowest order. Due to the form of the coefficient function, this
contribution is increasing almost up to the threshold, where it falls off
steeply.

The ratio between the light quark and gluon densities is twice as big in
Model~II as it is in Model~I. Therefore, we expect only a small
fraction of resolved events in Model~II to contain charm quarks,
and the main charm contribution will come from the direct fusion process.
The entirely different shapes of these contributions close to the
threshold explain  the difference between the charm  distributions in
the two models.

\begin{figure}
\begin{center}
{}~ \epsfig{file=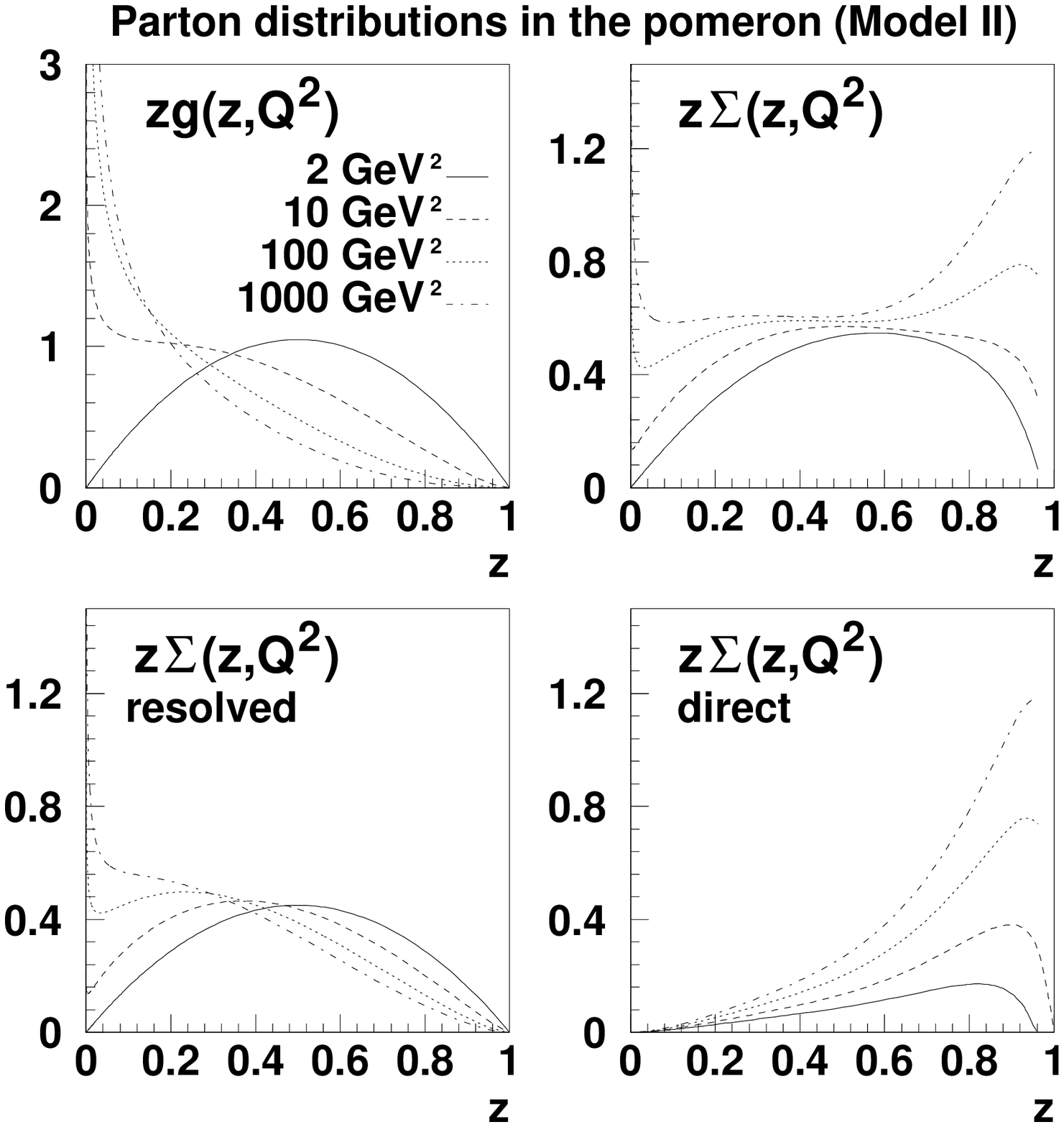,width=14cm}
\caption{Parton distributions in the pomeron, assuming a valence-like
structure of the resolved component at $Q_0^2=2\;\mbox{GeV}^2$ and a
pointlike coupling with strength $c=1$ of the
pomeron to quarks (Model~II). The relative normalizations between
resolved gluons and resolved light quarks are
chosen to be 70:30 at $Q_0^2$.
The lower plots show the resolved (Eq.~\protect{\ref{lightf22}}) and direct
(Eq.~\protect{\ref{ldf2}})  contributions to the light-quark singlet
distribution.}

\label{fig:partons2}
\end{center}
\end{figure}

Figure~\ref{fig:partons2} shows the parton distributions in Model~II,
where again exact SU(3) flavour symmetry is assumed.
The separate plots for the direct and resolved light-quark densities
show the origin of the high- and low-$z$ peaks in the quark singlet
distribution.

As both models give drastically different predictions for the
$Q^2$-evolution of the pomeron structure function in the high-$z$
region and for the kinematical distribution of charm quarks, we
expect that improved data will be able to distinguish between the
different approaches.

\section{Conclusions}
\label{sec:conc}

The idea that the pomeron has partonic structure
\cite{ing1} has been given strong support by the recent
measurements of the diffractive structure function at HERA.
In this paper we have presented a detailed study of
deep inelastic electron-pomeron scattering. We first derived
the complete set of kinematic variables for the deep inelastic
diffractive cross section. We showed that  when expressed  in terms
of appropriate variables this cross section is expected to factorize into
a pomeron structure function multiplied by a pomeron emission
factor, the latter being obtainable from hadron-hadron cross
sections.\footnote{Note that factorization is a `high-$Q^2$' phenomenon,
and  will of course break down in
the $Q^2 \to 0$ limit, see for example \cite{soper}.}
At present the variables which define the pomeron momentum are
not directly measured, although they can be inferred from the observed
hadronic final state. However, in terms of the measured variables
the factorization is only approximate. In Section~\ref{sec:kin}
we quantified the corresponding systematic error, and showed that it
was below the present level of experimental precision.

When the remnant protons are eventually detected at HERA, it should
be possible to measure their scattering angle relative to the electron
in the transverse plane. If the electron-pomeron scattering
picture is correct,
this distribution is predicted to be non-uniform, with a preference
for back-to-back scattering.  We presented  quantitative predictions
for this angular distribution in Sec.~\ref{sec:corr}, using the
Donnachie-Landshoff parametrization for the pomeron emission factor.

Finally, we presented two simple phenomenological models for the pomeron
structure function. The first is  based on the idea that
 at a `bound-state' $Q^2$ scale,
the pomeron consists predominantly of valence-like gluons, with a small
admixture of valence-like quarks. A momentum sum rule
is imposed. At higher $Q^2$-scales the distributions
are determined by standard GLAP perturbative evolution. Our starting
quark distributions are identical in shape, and similar in size,
to those calculated by Donnachie and Landshoff. In this model it is necessary
to rescale the pomeron flux factor (by a factor of approximately 2
in the case of the DL function) to account for the normalization
of the H1 data, whereupon good agreement is obtained with  the
measured $z$ and
$Q^2$ dependence of the pomeron structure function.
The light $(u,d,s)$
quarks carry about 17--25\% of the pomeron's momentum in the range
of $Q^2$ currently measured by H1.
Note that the fact that our quark and gluon distributions are `hard'
in the HERA $Q^2$ range means that standard linear GLAP evolution
should be perfectly adequate. Gluon-recombination effects,
giving rise to non-linear evolution (as studied for example in
Ref.~\cite{ing2}),  would  eventually be expected to become important
at very high $Q^2$ when the distributions have evolved to low $z$.

Our second model allows for a point-like coupling of the pomeron
to quarks, which generates an additional `direct' component for
the structure function, in analogy to the photon structure function.
In this model the pomeron flux factor obtained from soft hadronic processes
requires no further rescaling, since the direct component is large and
positive.
In contrast to the resolved (hadron-like) contribution,
the direct
contribution grows linearly with $\ln Q^2$, at all $z$. Thus
the asymptotic $Q^2 \to \infty $ behaviours of the two models
are very different, even though the current data are not yet
precise enough to
discriminate between them.

The experimentally-measured $(z,Q^2)$ range of the pomeron
structure function includes the charm quark threshold region. This
requires special treatment, since the charm contribution
to $F_2^P$ is expected to be significant above threshold.
Motivated by the successful treatment of the charm content of the proton,
 we have calculated this using the photon--gluon fusion process,
which takes the threshold kinematics correctly into account.
We have found  that the charm contribution to $F_2^P$ is indeed sizeable,
especially at high $Q^2$ and low $z$. Indeed, in Model~I the rapid increase of
the charm contribution with increasing $Q^2$ appears to account for
the bulk of the observed $Q^2$ dependence. In Model~II a similarly
large charm contribution is obtained, but this time mainly from the
direct coupling of the pomeron to charm quarks. The distinguishing
feature between the two models is that in Model~II the charm
contirbution is more concentrated at large $z$ (compare
 Figs.~\ref{fig:f2p1} and \ref{fig:f2p2}).

Our results on the quark and gluon content of the pomeron have many
implications. As already mentioned, we expect that a significant
fraction of hard diffractive scattering events will contain charm,
and our distributions provide a way of quantifying this.
Especially in  Model~I, the overall magnitude of the gluon distribution
compared to the quark
distribution
also predicts a large value for the pomeron's $R$-factor.
In particular, we expect $R^P \sim O(1)$,
in contrast to $R\sim O(\alpha_s) \ll 1$ for the proton, which results
in a similar magnitude of $R^{D\!S}$.
However, a consistent estimate of this would require a full
next-to-leading order perturbative calculation, which is beyond the
scope of the present paper.

In summary, we have shown that  simple quark and gluon
parton models of the pomeron, combined with  pomeron emission factors
extracted from soft hadronic processes, give an excellent
description of the H1 data. There are many ways in which this simple
picture can be tested, and our two models distinguished,
both at HERA and elsewhere. In the short term,
the measurement of the $\phi_{ep}$ correlation, the $Q^2$
dependence of the pomeron structure function
 particularly at large $z$,  and the
identification of the predicted
large charm contribution to the diffractive structure function
appear to offer the best possibilities.

\section*{Acknowledgements}

\noindent  Financial support from  the UK PPARC (WJS), and from
the Gottlieb Daimler- und Karl Benz-Stiftung and the
Studienstiftung des deutschen Volkes (TG) is gratefully acknowledged.
We thank Albert De Roeck, Peter Landshoff,
Peter Schlein and Juan Terron  for useful discussions.
This work was supported in part by the EU Programme
``Human Capital and Mobility'', Network ``Physics at High Energy
Colliders'', contract CHRX-CT93-0357 (DG 12 COMA).
\goodbreak


\newpage

\newpage

\newpage

\newpage

\newpage

\newpage

\newpage

\newpage

\begin{thebibliography}{10}

\bibitem{zeus}
ZEUS collaboration: M.~Derrick et al.,  Phys. Lett. {\bf B315} (1993) 481;
{\bf B332} (1994) 228; {\bf B338} (1994) 483.
\bibitem{h1}
H1 collaboration: T.~Ahmed et al., Nucl. Phys. {\bf B429} (1994) 477.
\bibitem{dldata}
A.~Donnachie and P.V.~Landshoff: Nucl. Phys. {\bf B244} (1984) 322;
{\bf B267} (1986) 690.
\bibitem{WA91}
WA91 collaboration: S.~Abatzis et al., Phys. Lett. {\bf B324} (1994) 509.
\bibitem{ing1}
G.~Ingelman and P. Schlein,  Phys. Lett. {\bf B152} (1985) 256.
\bibitem{UA8}
UA8 collaboration: R.~Bonino et al.,  Phys. Lett. {\bf B211} (1988) 239;
A.~Brandt et al., Phys. Lett. {\bf B297} (1992) 417.
\bibitem{SCHLEIN}
P.~Schlein, Proc. Int. Europhysics Conf. on High Energy Physics,
Marseille, France, July 1993, Editions Frontieres, eds. J.~Carr and
M.~Perrottet.
\bibitem{dldis}
A.~Donnachie and P.V.~Landshoff, Phys. Lett. {\bf B191} (1987) 309;
{\bf B198} (1987) 590(E).
\bibitem{ing2}
G.~Ingelman and K.~Janson-Prytz, Proc. Workshop `Physics at HERA',
eds. W.~Buchm\"uller and G.~Ingelman, October 1991, p.~233. \\
G.~Ingelman and K.~Prytz, Z. Phys. {\bf C58} (1993) 285.
\bibitem{dlpom}
P.V.~Landshoff and J.C.~Polkinghorne, Nucl. Phys. {\bf B32} (1971) 541. \\
G.A.~Jaroskiewicz and P.V.~Landshoff, Phys. Rev. {\bf D10} (1974) 170.
\bibitem{bfklph}
A.H.~Mueller and H.~Navelet, Nucl. Phys. {\bf B282} (1987) 727. \\
M.~Genovese, N.N.~Nikolaev and B.G.~Zakharov, KFA-J\"ulich preprint
KFA-IKP(Th)-1994-37 and references therein.
\bibitem{bfkl}
L.N.~Lipatov, Sov. J. Nucl. Phys. {\bf 23} (1976) 338.\\ E.A.~Kuraev,
L.N.~Lipatov and V.S.~Fadin, Sov. Phys. JETP {\bf 45} (1977) 199. \\
Ya.Ya.~Baltitsky and L.N.~Lipatov, Sov. J. Nucl. Phys. {\bf 28} (1978) 822.
\bibitem{dirac}
H.A.~Bethe and R.~Jackiw: {\it Intermediate Quantum Mechanics.}
Benjamin, Reading, Mass. (1968).
\bibitem{h1new}
H1 collaboration: T.~Ahmed et al., Phys. Lett. {\bf B348} (1995) 681.
\bibitem{debate}
H.~Fritzsch and K.H.~Streng, Phys. Lett. {\bf B164} (1985) 391. \\
N.~Arteaga-Romero, P.~Kessler and J.~Silva, Mod. Phys. Lett. {\bf A1}
 (1986) 211; {\bf A4}  (1989) 645.\\
E.L.~Berger, J.C.~Collins, D.E.~Soper and G.~Sterman,
Nucl. Phys. {\bf B286} (1987) 704. \\
A.~Donnachie and P.V.~Landshoff, Nucl. Phys. {\bf B303} (1988) 634.\\
K.H.~Streng, Proc. HERA Workshop, ed. R.D.~Peccei, DESY Hamburg (1988),
Vol.~1, p.~365. \\
J.~Bartels and G.~Ingelman, Phys. Lett. {\bf B235} (1990) 175.\\
A.~Donnachie, Proc. Int. Workshop on Deep Inelastic Scattering
and Related Subjects, Eilat, Israel, February 1994, University
of Manchester preprint  M-C-TH-94-07 (1994).
\bibitem{gluons}
F.E.~Low, Phys. Rev. {\bf D12} (1975) 163. \\
S.~Nussinov, Phys. Rev. Lett. {\bf 34} (1975) 1286; Phys. Rev. {\bf D14}
(1976) 246.
\bibitem{DLNPB}
A.~Donnachie and P.V.~Landshoff, Ref.~\cite{debate}.
\bibitem{CTEQ}
J.C.~Collins et al., Phys. Rev. {\bf D51} (1995) 3182.
\bibitem{GOULIANOS}
K.~Goulianos, Rockefeller University preprint RU 95/E-06 (1995).
\bibitem{KRAMER}
B.A.~Kniehl, H.G.~Kohrs abd G.~Kramer, Z. Phys. {\bf C65} (1995) 657.
\bibitem{CAPELLA}
A.~Capella, A.~Kaidalov, C.~Merino and J.~Tran Thanh Van, Phys. Lett.
{\bf B343} (1995) 403.
\bibitem{glap}
V.N.~Gribov and L.N.~Lipatov,
Sov. J. Nucl. Phys. {\bf 15} (1972) 438, 675. \\
G.~Altarelli and G.~Parisi, Nucl. Phys. {\bf B126} (1977) 298. \\
Yu.L.~Dokshitzer, Sov. Phys. JETP {\bf 46} (1977) 641.
\bibitem{dynamical}
G.~Altarelli, N.~Cabibbo, L.~Maiani and  R.~Petronzio, Nucl. Phys. {\bf
B69} (1974) 531. \\
 M.~Gl\"uck and  E.~Reya, Nucl. Phys. {\bf B130} (1977) 76.
\bibitem{grv}
M.~Gl\"uck, E.~Reya and A.~Vogt, Z. Phys. {\bf C48} (1990) 471, {\bf
C53} (1992) 127.
\bibitem{grvcharm}
M.~Gl\"uck, E.~Reya and M.~Stratmann, Nucl. Phys. {\bf B422} (1994) 37.
\bibitem{grvnew}
M.~Gl\"uck, E.~Reya and A.~Vogt, Dortmund preprint DO-TH 94/24 (1994).
\bibitem{CFS} L.~Frankfurt, M.~Strikman, Phys. Rev. Lett. {\bf 64}
(1989) 1914. \\
 J.C.~Collins, L.~Frankfurt, M.~Strikman, Phys. Lett. {\bf B307}
(1993) 161.
\bibitem{soper}
A.~Berera and D.E.~Soper, Phys. Rev. {\bf D50} (1994) 4328.
\bibitem{HERA}
H1 Collaboration: T.~Ahmed et al., DESY preprint 95-006.\\
ZEUS Collaboration: M.~Derrick et al., Z. Phys. {\bf C65} (1995) 379.
\bibitem{mrsg}
A.D.~Martin, R.G.~Roberts and W.J.~Stirling: Durham University
preprint DTP/95/14.
\bibitem{cteq3}
CTEQ collaboration: H.L.~Lai {\it et al.}, Michigan
State University preprint, MSU-HEP-41024 (1994).
\bibitem{newzeus}
ZEUS Collaboration: M.~Derrick et al., DESY preprint 95-093.

\end{thebibliography}
\end{document}